\documentclass[fleqn,usenatbib]{mnras}

\usepackage{newtxtext,newtxmath}
\usepackage{float}
\usepackage{newtxmath}
\usepackage{adjustbox} 
\usepackage{graphicx}
\usepackage{lscape}
\usepackage{wrapfig}
\usepackage{epstopdf}
\usepackage{subfigure}
\usepackage{epstopdf}
\DeclareGraphicsExtensions{.ps}
\usepackage{booktabs}
\usepackage{adjustbox} 
\usepackage[T1]{fontenc}
\usepackage{float} 
\usepackage{gensymb}

\DeclareRobustCommand{\VAN}[3]{#2}
\let\VANthebibliography\thebibliography
\def\thebibliography{\DeclareRobustCommand{\VAN}[3]{##3}\VANthebibliography}

\usepackage{graphicx}	% Including figure files
\usepackage{amsmath}	% Advanced maths commands
\usepackage[dvipsnames]{xcolor}

\title{X-ray variability of the HMXB Cen X-3: evidence for inhomogeneous accretion flows.}

\author[]{
Sanjurjo-Ferr\'{i}n, G.$^{1}$
Torrej\'on, J.M.$^{1}$%\thanks{E-mail: mn@ras.org.uk (KTS)}
Postnov, K.$^{2,4}$
Oskinova, L.$^{3}$ $^{4}$
Rodes-Roca, J.J.$^{1}$
Bernabeu, G.$^{1}$
\\
$^{1}$Instituto Universitario de F\'isca Aplicada a las Ciencias y las Tecnolog\'ias, Universidad de Alicante, 03690 Alicante, Spain\\
$^{2}$ Sternberg Astronomical Institute, Moscow M.V. Lomonosov State University, Universitetskij pr, 13, Moscow 119234, Russia\\
$^{3}$Institute for Physics and Astronomy, Universit\"{a}t Potsdam, 14476 Potsdam, Germany\\
$^{4}$Department of Astronomy, Kazan Federal University, Kremlevskaya Str 18,
Kazan, Russia
}

\date{Accepted XXX. Received YYY; in original form ZZZ}

\pubyear{2020}

\begin{document}
\label{firstpage}
\pagerange{\pageref{firstpage}--\pageref{lastpage}}
\maketitle

\begin{abstract}
Cen X-3 is a compact high mass X-ray binary likely powered by Roche lobe overflow. We present a phase-resolved X-ray spectral and timing analysis of two pointed {\it XMM-Newton} observations. The first one took place during a normal state of the source, when it has a luminosity $L_{\rm X}\sim 10^{36}$ erg s$^{-1}$. This observation covered orbital phases $\phi= 0.00-0.37$, i.e. the egress from the eclipse. The egress lightcurve is highly structured, showing distinctive intervals. We argue that different intervals correspond to the emergence of different emitting structures. The lightcurve analysis enables us to estimate the size of such structures around the compact star, the most conspicuous of which has a size $\sim 0.3R_{*}$, of the order of the Roche lobe radius. During the egress, the equivalent width of Fe emission lines, from highly ionized species, decreases as the X-ray continuum grows. On the other hand, the equivalent width of the Fe\,K$\alpha$ line, from near neutral Fe, strengthens. This line is likely formed due to the X-ray illumination of the accretion stream. The second observation was taken when the source was 10 times X-ray brighter and covered the orbital phases $\phi= 0.36-0.80$.  The X-ray lightcurve in the high state shows dips. These dips are not caused by absorption but can be due to instabilities in the accretion stream. The typical dip duration, of about 1000~s, is much longer than the timescale attributed to the accretion of the clumpy stellar wind of the massive donor star, but is similar to the viscous timescale at the inner radius of the accretion disk.

\end{abstract}

\begin{keywords}
stars:accretion disk, eclipse -- (Stars:) pulsars: individual Cen X-3, X-rays: binaries
\end{keywords}

%%%%%%%%%%%%%%%%%%%%%%%%%%%%%%%%%%%%%%%%%%%%%%%%%%

%%%%%%%%%%%%%%%%% BODY OF PAPER %%%%%%%%%%%%%%%%%%

\section{Introduction}

In high-mass X-ray binaries (HMXBs),  a compact object (a neutron star -NS- or a black hole) orbits a massive star (the companion) accreting matter from its powerful stellar wind. HMXBs are key astrophysical laboratories where the process of accretion, the structure of the accretion stream and the donor's stellar wind can be studied in detail \citep[][for a review]{2017SSRv..212...59M}. They are progenitors of double degenerate binary mergers powering gravitational wave sources \citep{2019arXiv190106939V}. Although the evolutionary paths leading to HMXBs has been known for a long time \citep{1972NPhS..239...67V, 1973NInfo..27...58T}, some of the basic predictions still await observational confirmation. When the massive donor evolves, it can start to fill its Roche lobe, initiating the mass transfer through the inner Lagrangian point. In the case of short orbital periods, the unstable mass transfer onto the compact object can result in the formation of the common envelope (CE) stage that may lead to the merging of the components. The subject of this study, Cen X-3, has an orbital period of only 2.1 d, and the study of the mass transfer onto the compact star in this source is important from the point of view of the close binary evolution.    

\begin{table*}
\caption {Properties of the Cen X-3 system.  }
\centering
\begin{tabular}{lcc}
\midrule\midrule\\
\multicolumn{3}{c}{Donor Star}\\
\midrule
    MK type & O6-8 III & \citet{1979ApJ...229.1079H}\\
    $M_{\rm opt} $ & $20.2^{+1.8}_{-1.5}$  $M_\odot$ & \citet{2007AA...473..523V} \\
    
    $R_\ast$ & $12.1 \pm 0.5 ~ {R_\odot}$ & \citet{2011ApJ...737...79N} \\
        $E(B-V)$  &       2.456 {\rm mag}             & \citet{2013MNRAS.431..394W}\\
%          &       &  \\
\midrule          
\multicolumn{3}{c}{Neutron Star}\\
\midrule
     $M_{\rm NS}$ & $1.34^{+0.16}_{-0.14}$  $M_\odot$  & \citet{2007AA...473..523V} \\
    Spin period & 4.82 s  &\citet{2007AA...473..523V} \\
   
     Magnetic field    & $(2.4-3.0) \times 10^{12}$  G &  \citet{2011ApJ...737...79N}\\
%            &       &  \\
\midrule            
\multicolumn{3}{c}{Orbit}\\ 
\midrule
    Orbital period  & 2.087113936(7) d  & \citet{falanga}\\
    $\dot P/ P$ & $-1.738(4) \times10^{-6}$  yr$^{-1}$ & \citet{2007AA...473..523V}\\
    $i$ & $72 \degree ^{+6}_{-5}$  & \citet{2007AA...473..523V} \\
    $i$ &$79  \pm 3$ \degree & This work \\
    Eccentricity $e$ & $<0.0016$  & \citet{1997ApJS..113..367B} \\
    Semimajor axis $a$& $1.58^{+0.05}_{-0.04}$  ~ $R_\ast$ & This work\\
    NS distance to the barycentre & $1.48 \pm 0.01$  ~ $R_\ast$ & This work\\
%    Orbital velocity &  $465^{+21}_{-18}$     km s$^{-1}$ & This work \\
    NS velocity with respect to the barycentre & $436 \pm 4$ km s$^{-1}$  & This work\\
    Distance & $8.7\pm 2.3  ~ {\rm kpc}$ & \citet{2018arXiv180611397T}\\
    $T_{0}$ (MJD) &  50506.788423(7)  &  \citet{falanga}  \\
  \hline
\end{tabular}

\label{parameters}

\end{table*}

Cen X-3 is an eclipsing X-ray pulsar, first observed by \citet{641ee8007ab8411c818bcc484f398d2f}, composed by a NS spinning with a period of $\sim$ 4.8~s and orbiting the giant O6-8 III counterpart V779 Cen \citep{1972ApJ...172L..79S, 1974ApJ...192L.135K, 1979ApJ...229.1079H}. The NS spin period was discovered with \textit{Uhuru} \citep{1972ApJ...172L..79S,1971ApJ...167L..67G}. 

%%%%%ciclotron
A cyclotron resonance scattering feature (CRSF) at $\sim$ 30 keV was detected with \textit{Ginga} \citep{1992ApJ...396..147N} and later confirmed by \textit{BeppoSAX} \citep{1998A&A...340L..55S} and \textit{RXTE/HEXTE} \citep{1999hxra.conf...25H}. In turn, \cite{article}  observes that the CRSF energy decreases, along the NS pulse, from 36 keV at the ascent down to 28 keV at the descent. This is explained by assuming an offset of the dipolar magnetic field with respect to the NS center.
%%%%%%%

The X-ray eclipse lasts $\sim 22\%$ of the orbit,  unveiling a short orbital period $\simeq$ 2.08 d. 
The optical star fills its Roche lobe. In such a situation, the formation of an accretion disk is very likely. The presence of a large accretion disk has indeed been claimed \citep{1986A&A...154...77T,1978ApJ...224..625P} and is consistent with the high luminosity of the source, the quasi periodic oscillations (QPOs) observed at 40 mHz \citep{1991PASJ...43L..43T,2008ApJ...685.1109R} and the observed NS spin-up trend of $\simeq$ 1.135 ms yr$^{-1}$, with some fluctuations \citep{1996ApJ...456..316T}. Also, some aperiodic high and low X-ray states, with $\sim 125 / 165$ d time-scales, have been reported \citep{1983ApJ...273..709P}. This long-term variability is attributed to the precession of the accretion disk. Other structures might be present, such as an accretion wake \citep{Suchy_2008}, making Cen X-3 one of the most complex binaries known. 
 
The stellar wind of the donor is partially photo-ionized by the X-ray emission, producing emission lines due to recombination \citep{1996ApJ...457..397A,2003ApJ...582..959W}. These lines strengths, relative to continuum, change with the orbital phase and should be specially enhanced during the eclipse, when the direct continuum produced by the NS is blocked by the optical counterpart. The wind is focused to the inner Lagrangian point and is then stored temporarily in an accretion structure (likely a disk) from where it proceeds smoothly towards the NS magnetic poles. Consequently, the X-ray lighcurve of Cen X-3 does not show the chaotic random fluctuations typical of direct wind accretors, like Vela X-1 or 4U1700-37 \citep[][]{2017SSRv..212...59M}. This makes Cen X-3 an ideal benchmark to characterize the intrinsic properties of the accretion flow. To this end, we perform in this paper a detailed phase-resolved analysis of two {\it XMM-Newton} pointed observations (Table \ref{xmmobs}). 

Both observations have been studied previously. \citet{2010RAA....10.1127D} analysed Cen X-3 observations from different instruments, including the {\it XMM-Newton} eclipse-egress observation (ID 0111010101, Tab. \ref{xmmobs}). They note that in super-orbital high states the eclipse egress and ingress are sharper and shorter while in super-orbital low states they are shallower and longer, concluding that different flux states are caused by  a varying degree of obscuration by the precessing accretion disk, without ruling out that a part of the variation could be due to changes in the mass accretion rate.  Both observations were also analysed by \citet{2012BASI...40..503N}, with the objective of understanding the variability of the iron lines during the eclipse-egress and the out-of-eclipse phases. These authors conclude that Fe\,K$\alpha$ line forms close to the NS whereas the highly inonised Fe species are probably produced in the photo-ionized wind of the companion star or in the accretion disk corona. \citet{Aftab_2019} also concludes that the Fe\,K$\alpha$ emitting region is close to the X-ray source, despite that the ionization state of Fe is expected to be high near the compact object because of its intense X-ray emission. This is possible if Fe atoms closer to the source are in a very dense optically-thick structure, such as an accretion disk or a dense accretion stream.

\begin{figure}
\includegraphics[trim={6cm 2cm 4cm 2cm},width=1.05\columnwidth]{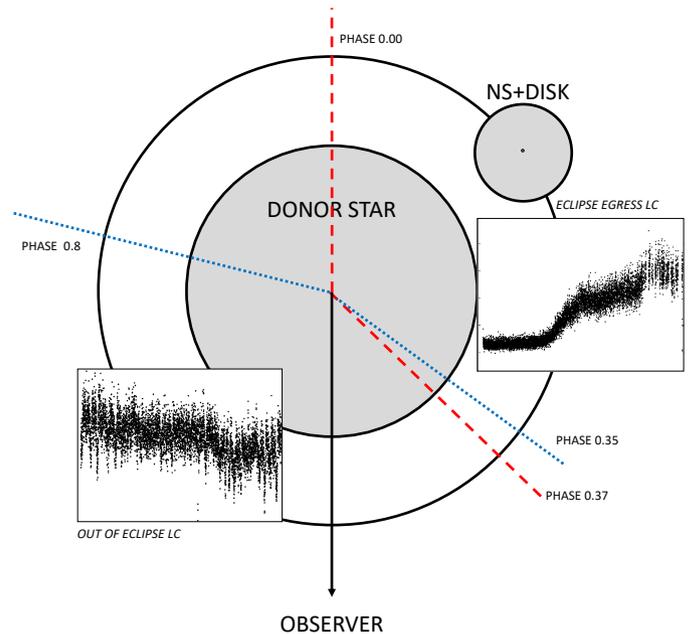}
\caption{Pole-on scaled sketch of the system and the orbital phases covered by the two {\it XMM-Newton} observations, obtained 5 years apart (Table \ref{xmmobs} and Fig. \ref{REXTE})  using the ephemeris of \citet{falanga}. The donor star radius, the orbit and the accretion disk are to scale. }
\label{systemsk}
\end{figure}

In Table \ref{parameters} we compile the system parameters relevant for this work, and in Fig. \ref{systemsk} we show a sketch of the system where the orbit, the donor and the putative disk (the size deduced from this work) are to scale.

 The paper is organized as follows. In Sect.2 we describe the observations and the analysis performed. In Sect. 3 we present our results. In Sect. 4 we discuss the results obtained and in Sect. 5 we summarize our findings.

%%%%%%%%%%%%%%%%%%%%%%%%%%%%%%%%%%%%%%%%
\section{Observations and analysis}

 We have analyzed two pointed observations performed with the X-ray Multi-Mirror Mission ({\it XMM-Newton}) space observatory. It carries three high throughput X-ray telescopes and one optical monitor. 

\begin{table}
\caption {\emph{XMM-Newton} Observation log. }
\centering
\begin{adjustbox}{max width=\columnwidth}
\begin{tabular}{ccccc}
\hline\hline
Observation ID & Date & Orbital phase & Duration & GTI duration\\
  &    &  &  (ks)& (ks)\\
\midrule
   0111010101 &  2001/01/27   &   0.00-0.37 &  68 & 57 \\
   0400550201  & 2006/06/12  & 0.35-0.80   & 81 & 79\\
  \hline

\end{tabular}
\end{adjustbox}
\label{xmmobs}
\end{table}

The log of observations is presented in Table \ref{xmmobs}. The first observation (ID 0111010101) was carried out using medium filters for the three European Photon Imaging Camera (EPIC) focal plane instruments, MOS1, MOS2 and pn. We did not use data from MOS1 camera as the instrument was operated in fast uncompressed mode. The MOS2 was operated in full frame mode and the pn was operated in small window mode. 
 This observation covers the eclipse egress and we will refer to it as egress in what follows (Fig. \ref{systemsk}). The data analyzed in the timing analysis (section \ref{res:sec:timming}) was taken with the pn camera in imaging mode. The data were first processed through the pipeline chains and filtered. For MOS2, only events with a pattern between 0 and 12  were considered. The data were filtered through \texttt{\#XMMEA EM}. For pn, we kept events with \texttt{flag = 0} and a pattern between 0 and 4. The data was filtered through \texttt{\#XMMEA EP} \citep{2001A&A...365L..27T}.  
In order to create the correspondent GTI (good time intervals) we chose a background threshold of < 0.35 counts s$^{-1}$ for MOS2 and < 0.4 counts s$^{-1}$ for pn. 
We also checked whether the observations were affected by pile-up using the task \texttt{epatplot} with negative results for the pn camera. The MOS2 camera showed pile-up, which was corrected by subtracting a small portion from the center of the extraction region, a circle centered in the brightest point of the source. The background was selected from a circle close to the source, avoiding bright pixels seen in the background. Finally, the resulting spectra for the pn and MOS2 cameras were combined into a single spectrum using the task \texttt{epicspeccombine}.

The second observation (ID 0400550201) was carried out by using thick filters for the focal plane instruments, MOS2 and pn. Both instruments were in timing mode, which attains the highest time resolution for {\it XMM-Newton}. The pn spectra were produced following the same steps as described above for the pn camera. In this case, the extraction region was a column centered in the brightest zone of the image while the background was selected from a column centered in a zone avoiding the light from the main source. We checked for the presence of pile-up with negative results. This observation took place totally out-of-eclipse and so will be named from now on. 

\begin{figure}
\centering
\includegraphics[trim={0cm 1cm 0cm 0cm},width=1\columnwidth]{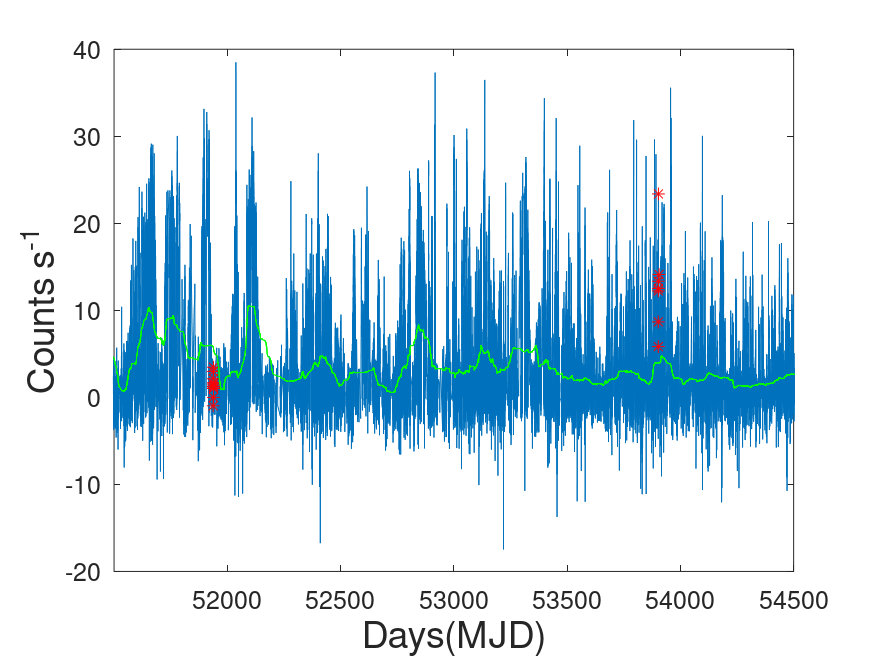}
\caption{The \textit{RXTE} long term lightcurve (over almost 20 years). Red asterisks show the periods when the {\it XMM-Newton} observations were performed. The green line represents the 500 bin moving average.}
\label{REXTE}
\end{figure}

During the egress the source had a luminosity of $L_{\rm X}\sim 10^{36}$ erg s$^{-1}$ while during the out-of-eclipse observation, the source was ten times brighter, $L_{\rm X}\sim 10^{37}$ erg s$^{-1}$. Both observations are covered by the \textit{RXTE-ASM} long term lightcurve (Fig. \ref{REXTE}). During the eclipse egress observation, the mean count rate is 1.4 $\pm$ 1.2 counts s$^{-1}$ while during the eclipse egress it is 13 $\pm$ 5 counts s$^{-1}$, close to the average of the top count rates displayed by the source, $\sim$ 16 $\pm$ 7 counts s$^{-1}$. Thus, the egress and out-of-eclipse observations will also be referred to as low state and high state, respectively.

The spectra were analyzed and modelled with the \textsc{xspec}\footnote{maintained by HEASARC at NASA/GSFC.} package. The energy range used for spectral fitting was 0.3$-$10 keV. The spectra were produced with a spectral bin size of 5 and not further modified in the analysis. The errors were obtained with the \texttt{error} task, provided by \textsc{xspec}, for a 90$\%$ confidence level. The emission lines were identified thanks to the \textsc{atomdb}\footnote{http://www.atomdb.org/} data base. 

The lightcurve timing analysis was performed only for the pn data because of its higher time resolution. The photon arrival times were transformed to the solar system barycentre. To analyze the lightcurves we used the \texttt{period} task inside the \textit{Starlink} suite\footnote{http://starlink.eao.hawaii.edu/starlink}. The \textsc{period04} program was also used \citep{2005CoAst.146...53L}. The \textsc{period04} package is especially suited for the statistical analysis of large astronomical time series containing gaps\footnote{https://www.univie.ac.at/tops/Period04/}.

\begin{figure*}
\centering
\subfigure{\includegraphics[trim={2cm 0cm 2cm 2cm},width=0.85\textwidth]{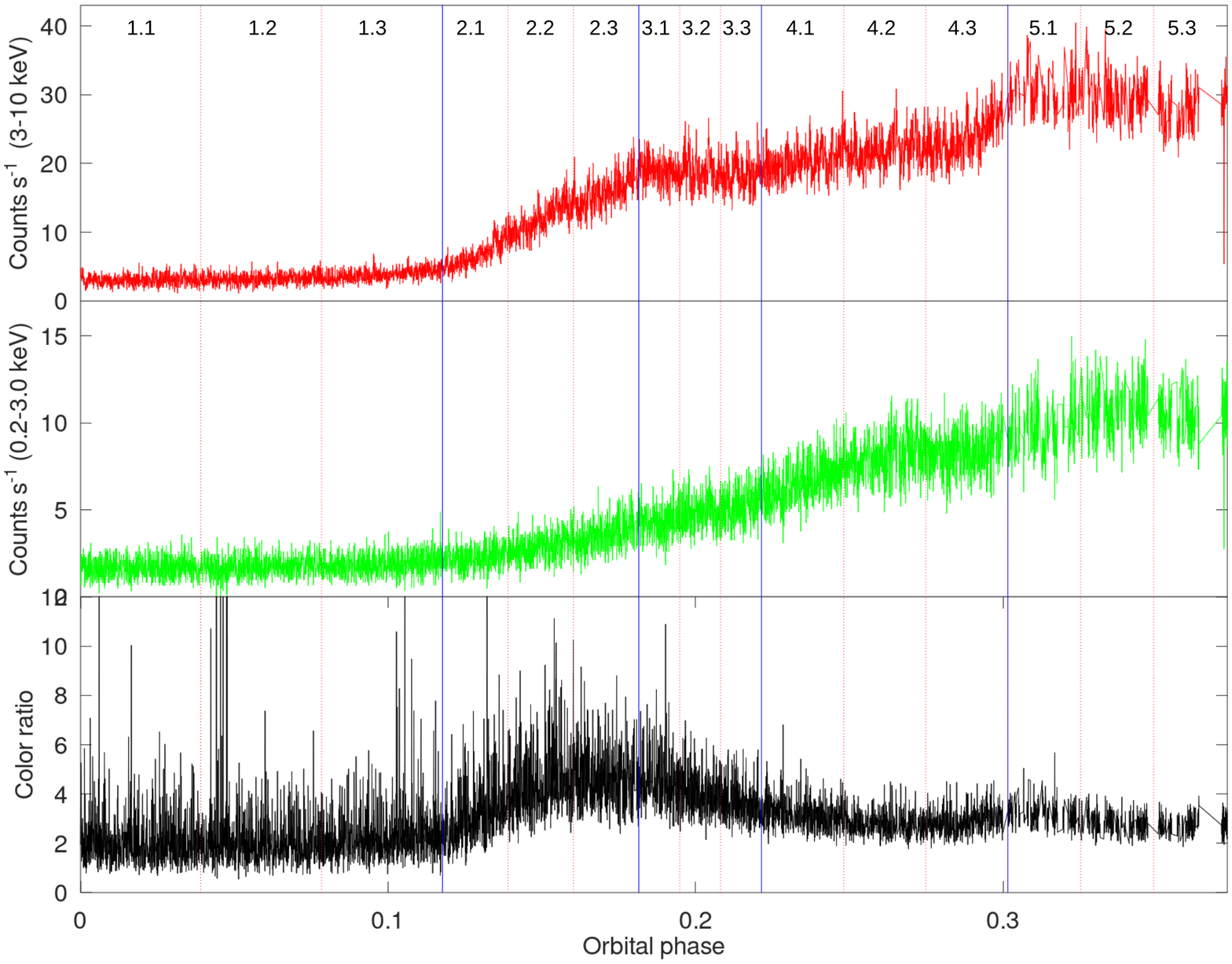}}
\subfigure{\includegraphics[trim={2cm 2cm 2cm 1.1cm},width=0.85\textwidth]{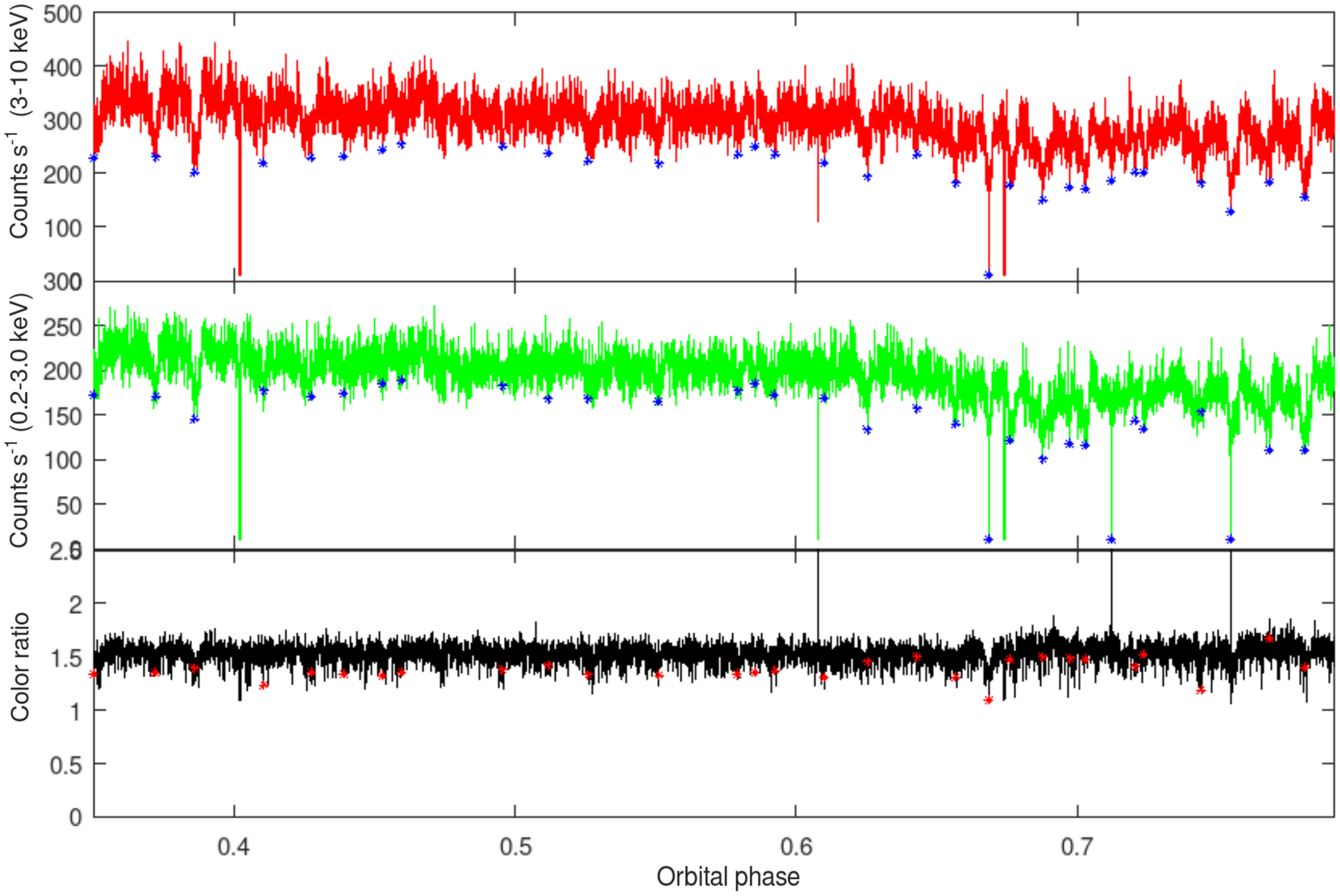}}
\caption{\textit{XMM-Newton} X-ray lightcurves of Cen X-3. The top three panels show the egress lightcurve: $3-10$ keV (red), $0.2-3$ keV (green), Color ratio (CR, black) defined as the ratio of lightcurves ($3-10$) keV/($0.2-3$) keV. The solid blue vertical lines show the five different intervals of the egress. Each interval was further subdivided into three intervals (dotted red vertical lines) of the same duration. The three bottom panels show the out-of-eclipse observation, taken five years later. This observation was divided into twenty different intervals of the same duration. Dips are marked by asterisks.} 
\label{lcurve}
\end{figure*}

%%%%%%%%%%%%%%%%%%%%%%%%%%%%%%%%%%%%%%%%%%%%%%%%%
\section{Results}

\subsection{XMM-Newton Timing}
\label{res:sec:timming}

\textit{The egress.} This observation took place during a low state of the source \citep{2010RAA....10.1127D}. The lightcurve shows a distinctive structure as the NS emerges from eclipse (Fig. \ref{lcurve}, upper panel). We have divided it in five different intervals. Roughly, there are three plateaus (denoted 1,3 and 5), at increasing flux levels, with duration of $\sim$ 21, 19 and 13 ks respectively. These plateaus are separated by two rises of different duration: 12 ks the first (2) and 14 ks the second (4). Interval 4 ends with a sudden, sharp rise lasting for 3.4 ks (interval 4.3). The first plateau (1) corresponds to the total eclipse.

The color ratio CR, defined here as the ratio of lightcurves (3$-$10) keV/(0.2$-$3) keV, changes during the egress. It shows a broad maximum during intervals 2 and 3, corresponding to the first rise. As will be shown by the spectral analysis, this is due to an increased absorption. After that (intervals 4 and 5), it remains more or less constant and slightly higher than at the beginning (the eclipse). 

\begin{figure}
\centering
\includegraphics[trim={2cm 0cm 1cm 1cm},width=1\columnwidth]{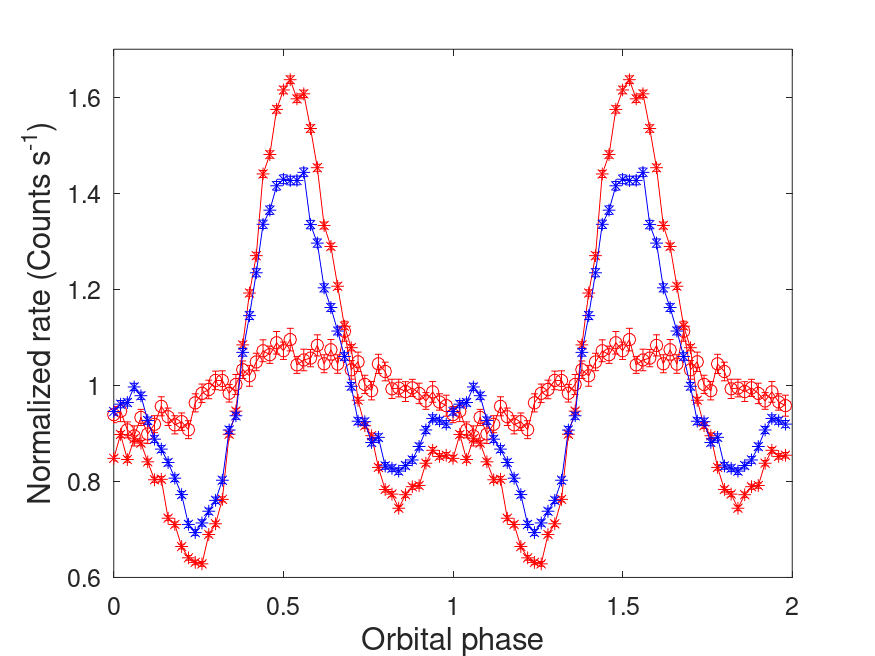}
\caption{Folded NS pulse of the egress (three last bins) in the high energy range ($3-10$ keV, red circles) and of the out-of-eclipse (first bin) in the low-energy range (0.3$-$3.0 keV, blue asterisks) and the high-energy range (3$-$10 keV, red asterisks). These pulses took place at similar orbital phases ($\sim$ 0.32 and $\sim$ 0.36 respectively). The rate (counts\,s$^{-1}$) was normalized.}
\label{ecpulse}
\end{figure}

The NS pulsations are only detected in the last interval of the lightcurve (from orbital phase $\sim$ 0.25 onwards), once the NS is completely out of the eclipse. The high energy ($3-10$ keV) folded pulse is represented in Fig. \ref{ecpulse} (red circles). It demonstrates single broad peak. The pulsed fraction is much lower than that found out-of-eclipse (see below). No pulse is found at low energies ($0.2-3$ keV). 

\textit{The out-of-eclipse}. This observation was taken in a high state of the source. The \textit{XMM-Newton} lightcurve is rather stable, decreasing by $\sim$ 20$\%$ in brightness towards the end. 

\begin{figure}
\includegraphics[trim={3cm 2cm 3cm 1cm},width=1\columnwidth]{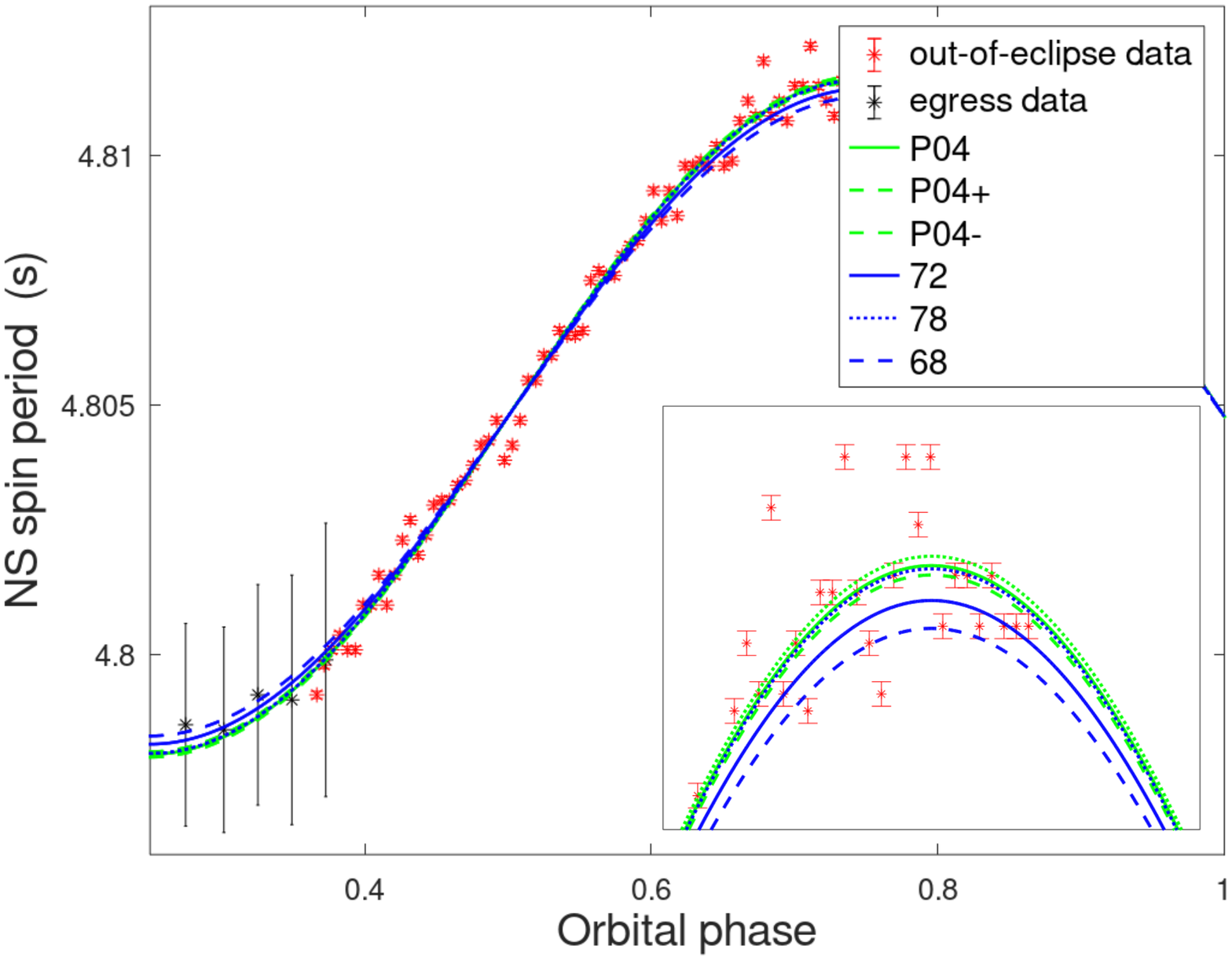}
\caption{Observed NS spin period evolution due to Doppler effect (red and black asterisks for eclipse and egress, respectively.). The green line is the predicted NS evolution calculated with \textsc{period04} and the dotted and dashed green lines represent the uncertainty. Error bars are of the order of the symbol size. The blue line and the dotted and dashed blue lines represent the theoretical Doppler effect calculated with the inclination given by \citet{2007AA...473..523V} and its uncertainty. The source distance and orbital radius were fixed at the values given in Table \ref{parameters}.}
\label{doppler}
\end{figure}

The NS spin pulsations are now clearly detected (Fig. \ref{ecpulse}) both at high and low energies (red and blue stars, respectively\footnote{only the first time bin is shown}). The pulse is now double peaked and shows a pulsed fraction much higher than during the low state (egress). To search for the NS pulse, we divided the lightcurve into 81 intervals of 1000 s each. The resultant NS spin period evolution is shown in Fig. \ref{doppler}. There is a clear spin down, caused by the Doppler effect of the NS orbiting the donor. 

The amplitude of the NS spin evolution can be used to calculate the system inclination. Ideally, this would require simultaneous radial velocity curves for the optical donor. Since this is not available here, we use {\it XMM-Newton} data to constrain its value. For a pulsar with period $P_0$ on an orbit with an inclination angle $i$ and a radial velocity of the NS with respect to the system barycentre $v_{\rm D}$, the observed Doppler-shifted spin period, $P_{\rm D}$ is,

\begin{equation}
\label{Doppler}
\begin{split}
P_{\rm D}=P_{0}\left(1+\frac{v_{\rm D}}{c}\right)\\
v_{\rm D}=-r\omega \sin\phi \sin i\\
\end{split}
\end{equation} 

In Fig. \ref{doppler} we show the simulated Doppler effect for several system inclinations (solid lines) along with the observed data. Our results constrain this value to $i=79\pm3^{\circ}$, fully compatible with the range given by \citet{2007AA...473..523V}. 

Although the out-of-eclipse lightcurve is rather stable, a remarkable feature is the presence of multiple dips, i.e., short time intervals where the flux drops significantly below average. The presence of dips was previously reported by \citet{2011ApJ...737...79N} from the \textit{Suzaku} observations.  

\begin{figure}
\centering
\includegraphics[trim={2cm 1.7cm 2cm 1cm},width=1\columnwidth]{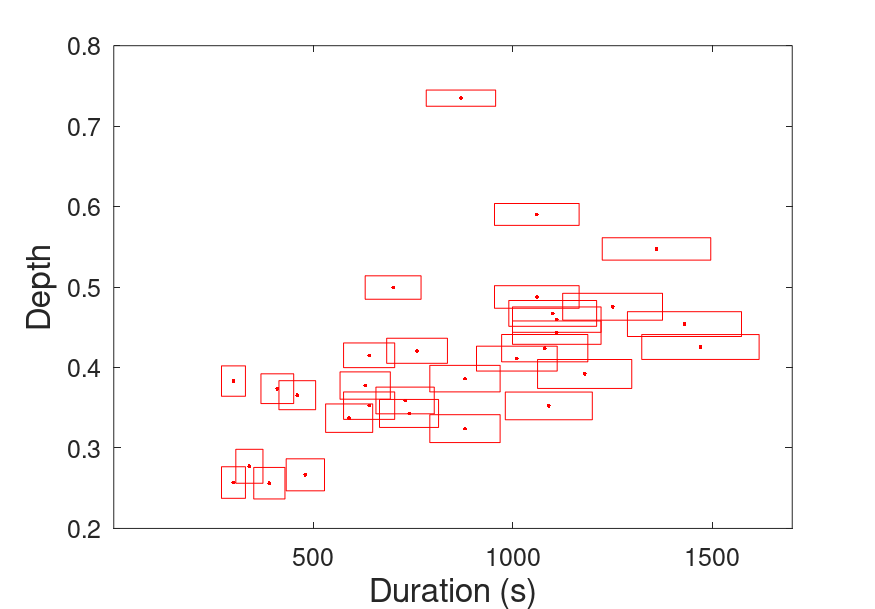}
\caption{Dip depth vs dip duration.}
\label{sq}
\end{figure}

In order to characterize the dips, we have implemented an automatic detection algorithm using pairs of moving averages and used it on the high-energy lightcurve. A decrease in counts will be classified as a dip if: a) the fast moving average counts go below the slowly moving average counts and b) if the maximum difference in counts between the slowly and the fast moving averages, at the dip position (the lower number of counts), is larger than 3 times the quadratic sum of the error of both moving averages. In this way, we avoid counting as dips random fluctuations of the lightcurve while keeping a good sensitivity. The beginning and the end of each dip are defined as the counts recovery to the slow moving average in the out-of-dip region. We also define the dip depth as the decrease in the number of counts divided by the number of counts of the slow moving average in the adjacent out-of-dip regions. About 30 dips have been so detected (Fig. \ref{lcurve}, blue asterisks). The results are summarized in Table \ref{dips}. Longer dips tend to be also deeper (Fig. \ref{sq}) and vice versa. The flux decreases by 40\%, on average, during the dips. However, the color ratio either remains unchanged or decreases (Fig. \ref{lcurve}, lower panel, red asterisks).

\begin{table*}
 \caption{Summary of the dips seen in the out-of-eclipse observation.}
 \begin{adjustbox}{max width=\textwidth}
  \centering
    \begin{tabular}{cccccccccc}
\hline\hline
    Phase & Duration & Depth & Color ratio  & Flux  &  Phase & Duration &Depth & Color ratio & Flux  \\

	   & (s)  &   &   &  ($\times 10^{-10}$ erg cm$^{-2}$ s$^{-1}$) & &(s) &   &   &  ($\times 10^{-10}$ erg cm$^{-2}$ s$^{-1}$) \\ 
	  \hline
0.35	&	$600\pm60$	&	$0.42\pm0.02$	&	$1.34\pm0.04$	&	$4.6_{-0.8}^{+1.1}$	&	0.63	&	$1400\pm140$	&	$0.45\pm0.02$	&	$1.45\pm0.05$	&	$4.8_{-0.6}^{+0.7}$	\\\\
0.37	&	$1000\pm100$	&	$0.41\pm0.02$	&	$1.35\pm0.04$	&	$4.8_{-0.8}^{+1.1}$	&	0.64	&	$500\pm50$	&	$0.27\pm0.02$	&	$1.5\pm0.05$	&	$4.4_{-0.8}^{+1.1}$	\\\\
0.39	&	$1100\pm110$	&	$0.49\pm0.01$	&	$1.39\pm0.05$	&	$5.0_{-0.7}^{+0.9}$	&	0.66	&	$1100\pm110$	&	$0.46\pm0.02$	&	$1.31\pm0.05$	&	$4.2_{-0.6}^{+0.8}$	\\\\
0.41	&	$1500\pm150$	&	$0.43\pm0.02$	&	$1.24\pm0.04$	&	$4.8_{-0.7}^{+0.8}$	&	0.67	&	$900\pm90$	&	$0.73\pm0.01$	&	$1.41\pm0.07$	&	$4.1_{-0.7}^{+0.9}$	\\\\
0.43	&	$1100\pm110$	&	$0.44\pm0.01$	&	$1.35\pm0.04$	&	$4.8_{-0.6}^{+0.8}$	&	0.68	&	$700\pm70$	&	$0.5\pm0.01$	&	$1.47\pm0.05$	&	$3.8_{-0.6}^{+0.9}$	\\\\
0.44	&	$1100\pm110$	&	$0.35\pm0.02$	&	$1.33\pm0.04$	&	$5.0_{-0.5}^{+0.7}$	&	0.69	&	$1300\pm130$	&	$0.48\pm0.02$	&	$1.5\pm0.06$	&	$3.7_{-0.8}^{+1.1}$	\\\\
0.45	&	$700\pm70$	&	$0.36\pm0.02$	&	$1.32\pm0.04$	&	$5.0_{-0.6}^{+0.8}$	&	0.7	&	$300\pm30$	&	$0.38\pm0.02$	&	$1.48\pm0.06$	&	$4.0_{-1.0}^{+1.7}$	\\\\
0.46	&	$900\pm90$	&	$0.32\pm0.02$	&	$1.35\pm0.04$	&	$5.2_{-0.8}^{+1.0}$	&	0.7	&	$1100\pm110$	&	$0.47\pm0.02$	&	$1.47\pm0.06$	&	$4.1_{-0.8}^{+1.0}$	\\\\
0.5	&	$400\pm40$	&	$0.26\pm0.02$	&	$1.38\pm0.04$	&	$4.7_{-0.9}^{+1.3}$	&	0.71	&	$400\pm40$	&	$0.37\pm0.02$	&	$2.54\pm0.11$	&	$4.3_{-0.9}^{+1.4}$	\\\\
0.51	&	$600\pm60$	&	$0.35\pm0.02$	&	$1.42\pm0.05$	&	$4.7_{-0.6}^{+0.8}$	&	0.72	&	$300\pm30$	&	$0.28\pm0.02$	&	$1.41\pm0.05$	&	$4.3_{-0.8}^{+1.2}$	\\\\
0.53	&	$900\pm90$	&	$0.39\pm0.02$	&	$1.33\pm0.04$	&	$4.6_{-0.7}^{+0.9}$	&	0.72	&	$500\pm50$	&	$0.37\pm0.02$	&	$1.52\pm0.05$	&	$4.1_{-1.0}^{+1.6}$	\\\\
0.55	&	$800\pm80$	&	$0.42\pm0.02$	&	$1.33\pm0.04$	&	$4.6_{-0.6}^{+0.7}$	&	0.74	&	$1100\pm110$	&	$0.42\pm0.02$	&	$1.19\pm0.04$	&	$4.1_{-0.6}^{+0.7}$	\\\\
0.58	&	$600\pm60$	&	$0.34\pm0.02$	&	$1.33\pm0.04$	&	$4.6_{-0.8}^{+1.1}$	&	0.75	&	$1100\pm110$	&	$0.59\pm0.01$	&	$1.44\pm0.06$	&	$3.8_{-0.6}^{+0.8}$	\\\\
0.59	&	$300\pm30$	&	$0.26\pm0.02$	&	$1.35\pm0.04$	&	$4.7_{-0.9}^{+1.4}$	&	0.77	&	$1200\pm120$	&	$0.39\pm0.02$	&	$1.67\pm0.06$	&	$4.1_{-0.8}^{+1.1}$	\\\\
0.59	&	$700\pm70$	&	$0.34\pm0.02$	&	$1.37\pm0.04$	&	$4.7_{-0.7}^{+0.9}$	&	0.78	&	$1400\pm140$	&	$0.55\pm0.01$	&	$1.4\pm0.06$	&	$4.1_{-0.5}^{+0.7}$	\\\\
0.61	&	$600\pm60$	&	$0.38\pm0.02$	&	$1.30\pm0.04$	&	$4.8_{-0.9}^{+1.2}$	&		&		&		&		&	\\

\hline
    \end{tabular}
    \end{adjustbox}
  \label{dips}%
\end{table*}%

\subsection{\textit{XMM-Newton} spectra}
\label{res:sec:spec}

\begin{figure*}
\centering
\subfigure{\includegraphics[trim={6cm 0cm 6cm 1cm},width=1\columnwidth]{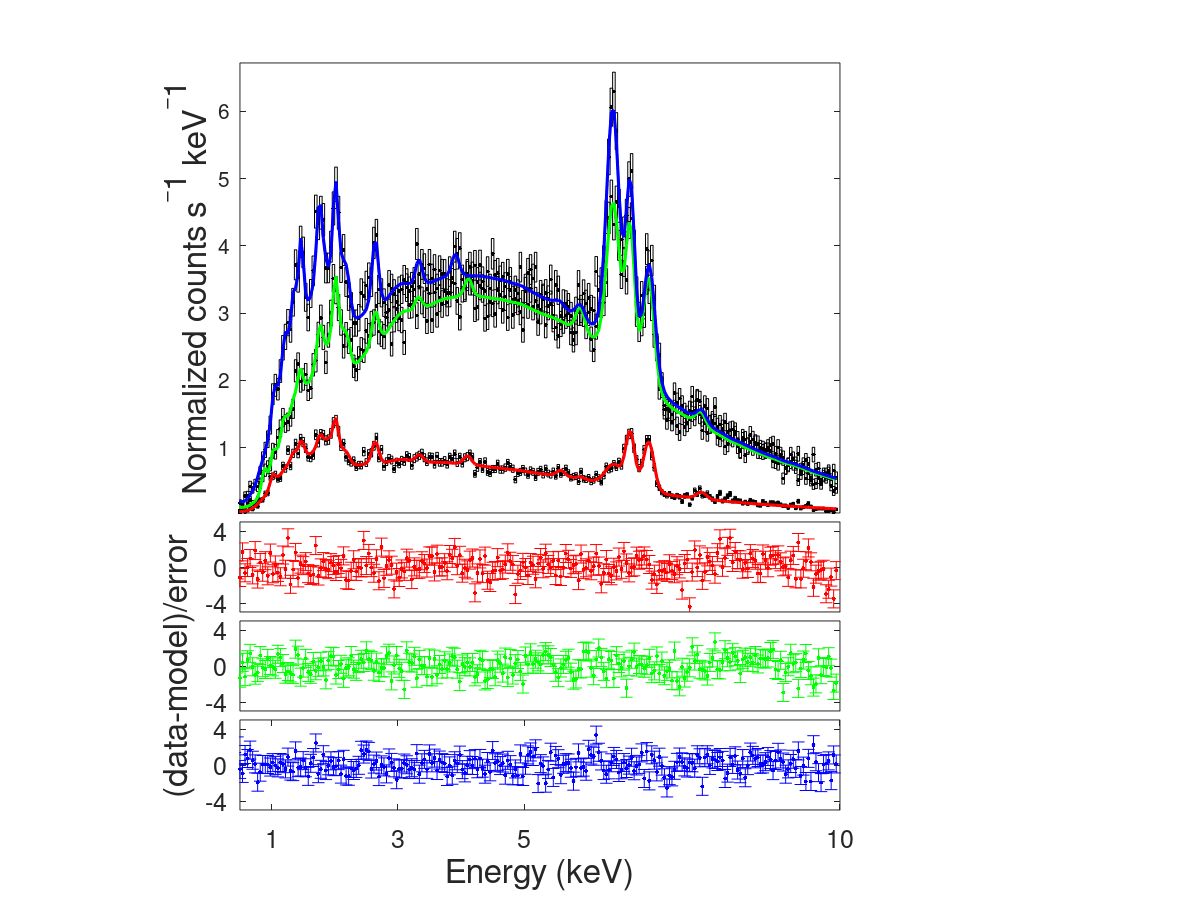}}
\subfigure{\includegraphics[trim={6cm 0cm 6cm 1cm},width=1\columnwidth]{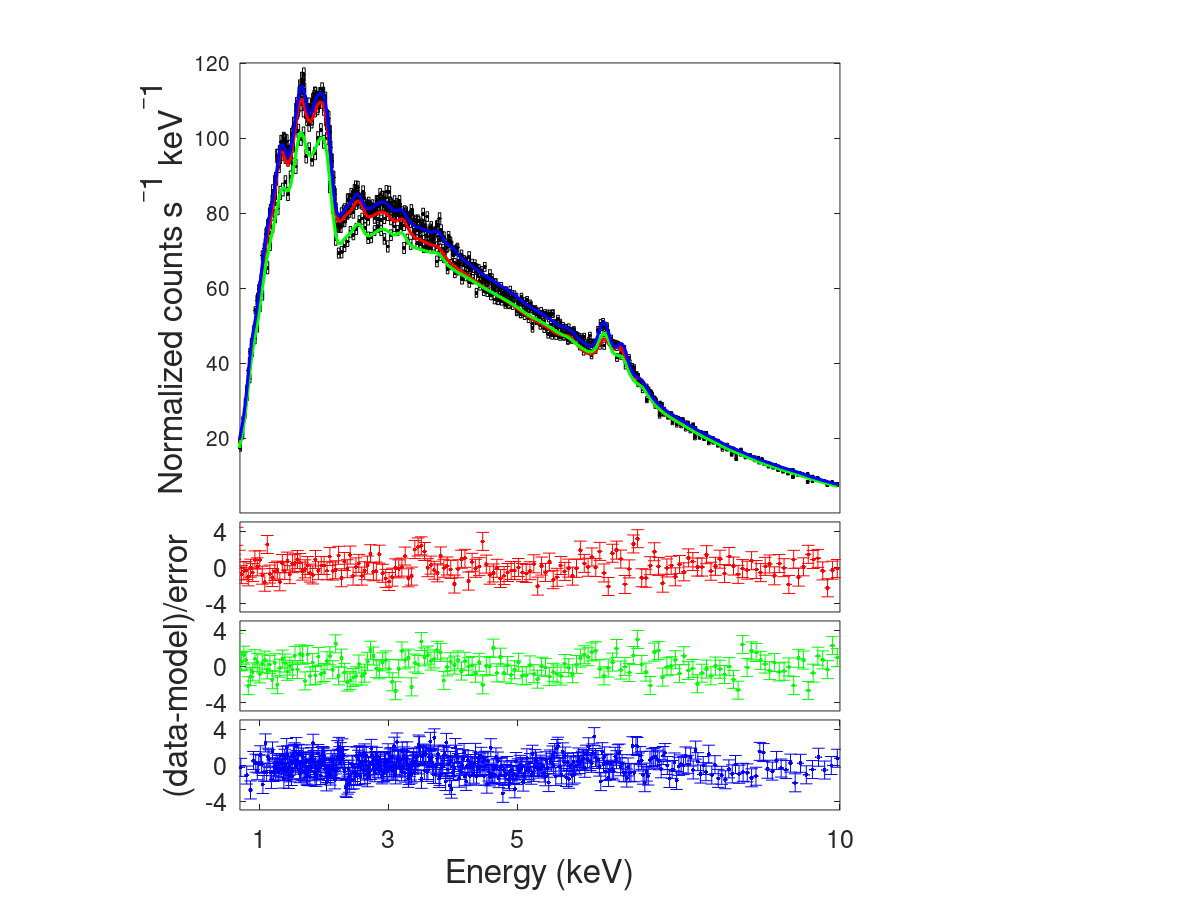}}
\caption{Selected spectra from the egress observation (left), intervals 1.3 (red), 3.3 (green) and 5.3 (blue ) and from the out-of-eclipse observation (right), dip 10 (blue), out-of-dip 26 (green) and dip 3 (red). The black squares represent the spectral data and uncertainties and the solid lines show the total model. Lower panels show the residuals of each fit in its corresponding color.}
\label{spec}

\subfigure{\includegraphics[trim={0cm 1.5cm 0cm 1cm},width=0.9\textwidth]{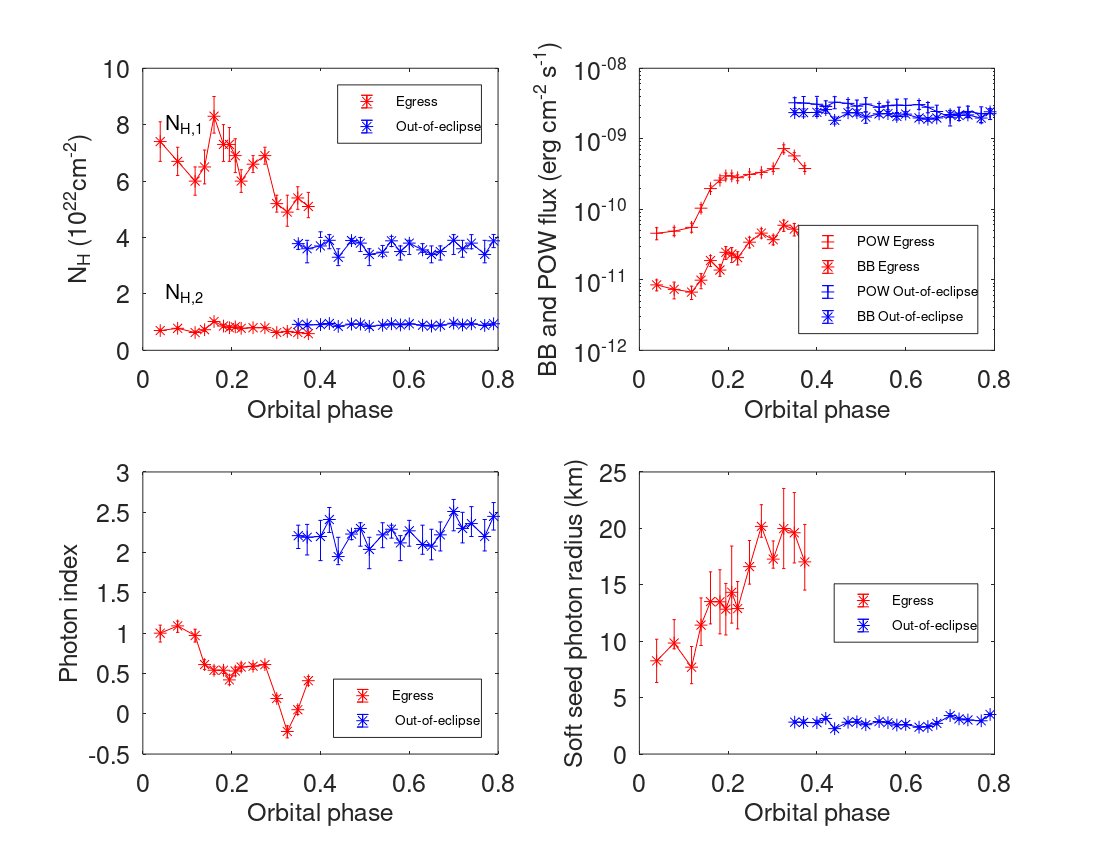}}

\caption {Evolution of some important parameters of the model. Red and blue symbols represent the egress and the out-of-eclipse data respectively. Top left panel shows the evolution of the absorption column. $N_{\rm H,2}$ is compatible with the ISM medium towards the system while $N_{\rm H,1}$ represents the ISM plus the local absorption. The right upper panel shows the evolution of the powerlaw and black body unabsorbed fluxes (note the logarithmic vertical axis). Bottom left panel shows the photon index $\Gamma$ of the powerlaw component. In the bottom right panel we show  the evolution of the soft seed photon radius, calculated from the temperature and luminosity.  }
\label{disk}
\end{figure*}

We performed a phase-resolved spectral analysis of both \textit{XMM-Newton} observations. The egress observation was initially divided into five different intervals described above (Fig. \ref{lcurve}). Due to the high variability, we further subdivided each part into three intervals obtaining, finally, fifteen spectra. For the out-of-eclipse observation we performed two different analyses. On one hand we divided the observation into twenty different intervals of the same duration, to explore the spectral evolution with orbital phase. On the other hand, to study the nature of dips, we performed a separate spectral analysis of each dip detected as well as of all the out-of-dip regions between them.

One spectral model was used to describe all spectra. Following \citet{Aftab_2019}, the best fit was achieved by combining a black body (\texttt{bbody} in \textsc{xspec)} plus a \texttt{powerlaw}. The success of this phenomenological model prompted us to try self-consistent comptonisation models but all of them gave poorer statistics ($\chi^2_{\rm r}\sim$ 1.2, on average, vs. 1.08 for the phenomenological model) as well as parameters difficult to interpret. For example, \texttt{compmag}, a hybrid model used to describe accretion onto a magnetized NS, yielded a temperature of the seed black body spectrum $kT_{\rm bb} =$ 2.2 keV, only barely lower than that of the comptonising cloud, $kT_{\rm e} =$ 2.8 keV (see also \cite{2016A&A...591A..29F} for the difficulties when fitting Cen X-3 \textit{Suzaku} and \textit{NuStar} data with \texttt{compmag}). Replacing the blackbody by other thermal components leads also to worse results. For example, \texttt{diskpbb} gave too high temperatures and also pointed to a significant radial advection, which is not expected in NS systems. Using \texttt{diskbb} or \texttt{disk} gives an internal disk radius smaller than, or comparable to, the NS radius. Eliminating the powerlaw component, resulted in a statistically acceptable fit but the soft photon source radius was unphysically small and the temperature was too high ($kT\sim 10-20$ keV).

The \texttt{bbody} model parameters include the temperature ($kT_{\rm bb}$) and the normalization, defined as $L_{39} D_{10}^{-2}$ where $D_{10}$ is the distance to the source in units of 10 kpc and $L_{39}$ is the luminosity in units of $10^{39}$erg s$^{-1}$. The \texttt{powerlaw} component is a simple photon powerlaw, where $\Gamma$ is a dimensionless photon index and the normalization, $K$, is the flux density in photons keV$^{-1}$ cm$^{-2}$ s$^{-1}$ at 1 keV. 

Besides the interstellar medium (ISM) absorption component, we also allowed for the presence of a local absorber, modulated by a partial covering fraction (parameter \textit{C}) which acts as a proxy for the degree of clumping in the stellar wind of the donor star. The ISM absorption is modeled by the X-ray absorption model T\"{u}bingen-Boulder \texttt{TBnew}\footnote{http://pulsar.sternwarte.uni-erlangen.de/wilms/research/tbabs/}. This model calculates the cross section for X-ray absorption by the ISM as the sum of the cross sections due to the gas-phase, the grain-phase, and the molecules in the ISM \citep{2000ApJ...542..914W}. 

The model used is described by Eq. \ref{model}.

\begin{equation}
\label{model}
\begin{split}
F(E) & = [C\exp(-N_{\rm H,1}\sigma(E))+\\
& +(1-C)\exp\left(-N_{\rm H,2}\sigma(E) \right)] \left[\texttt{bbody + powerlaw + G}\right ]
\end{split}
\end{equation} 

where $G$ represents the Gaussian functions added to account for the emission lines. 

As expected, the egress observation shows a progressively changing spectra. In turn, during the out-of-eclipse observation the spectra are found to be stable. In Fig. \ref{spec} some selected spectra are presented along with the model and residuals. The corresponding fit parameters are reported in the appendix, Tables \ref{appendix:A} (egress) and \ref{appendix:B} (out-of-eclipse). Figure \ref{disk} shows the evolution of some key model parameters. 

\textit{Emission lines.} The donor's stellar wind is photo-ionized by the NS X-rays, producing emission lines due to recombination \citep{1996ApJ...457..397A}. These lines are specially enhanced during the egress observation, when the direct emission from the NS is eclipsed by the companion, as was first observed by {\it ASCA}  \citep{8153132}.

\begin{figure}
\centering
\subfigure{\includegraphics[trim={1.5cm 1cm 17cm 0cm},width=1\columnwidth]{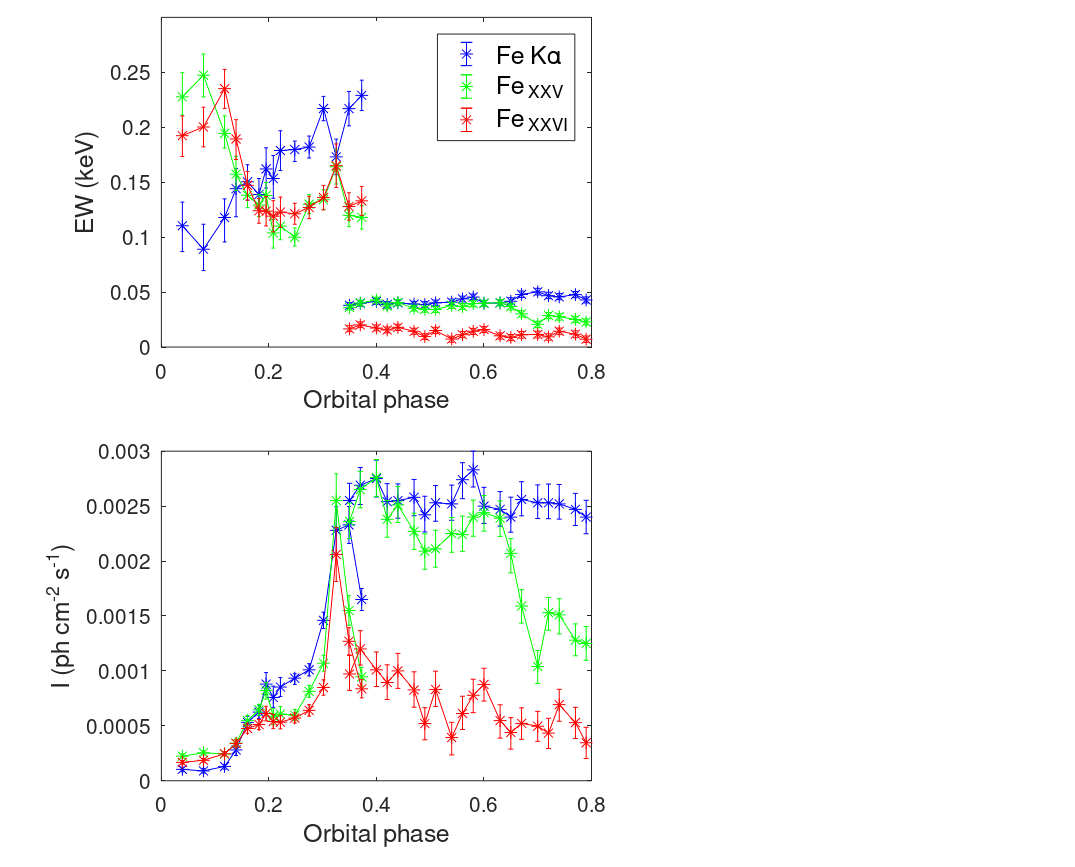}}
\caption{Evolution of the equivalent width (upper panel) and gaussian intensity (bottom panel) of the Fe emission lines with the orbital phase (Fe\,K$\alpha$ blue, \ion{Fe}{xxv} green, \ion{Fe}{xxvi} red). }
\label{felines}
\end{figure}

The most prominent lines in the EPIC spectra of Cen X-3 correspond to Fe (Fig. \ref{felines}). The equivalent width (EW, upper panel) of the highly ionized species, \ion{Fe}{xxv} (green) and \ion{Fe}{xxvi} (red) decrease, overall, as the NS emerges from eclipse. Surprisingly, the Fe\,K$\alpha$ line, from near neutral Fe (blue), strengthens. This clearly points to a different origin location. Once out-of-eclipse, the EW remains approximately constant. The lower panel depicts the line intensities for all three species. They all grow, overall, upon the eclipse egress. This behaviour is entirely consistent with that reported by \citet{2012BASI...40..503N}. When interpreting these plots, it should be taken into account that the state of the source was different, 10 times brighter at the orbital phases $\phi > 0.35$. 

\begin{figure*}
\centering
\subfigure{\includegraphics[trim={3cm 9cm 3cm 2cm},width=0.9\textwidth]{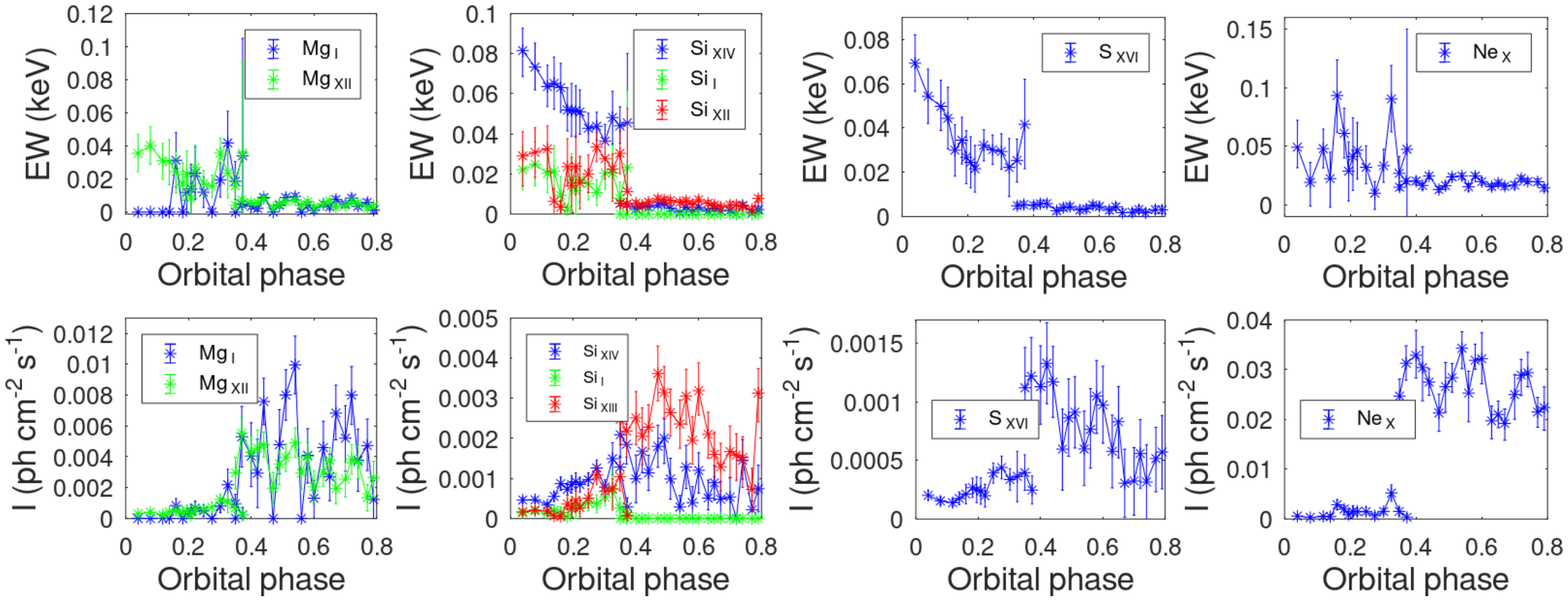}}
\caption{Gaussian intensity and equivalent width evolution of emission lines from Mg, Si, S and Ne, found in the analysis. The equivalent width of \ion{Si}{xiv} and \ion{S}{xvi} emission lines show a similar behaviour to the \ion{Fe}{xxv} and \ion{Fe}{xxvi} emission lines. The intensity of \ion{Mg}{i} and \ion{Si}{xiv} shows a maximum during phase 0.5.} 
\label{other_lines}
\end{figure*}

Prominent lines from Mg, Si, S and Ne are also seen in the spectra. The evolution of their intensity and EW can be seen in Fig. \ref{other_lines}. Although the associated errors are generally larger, they show an interesting behaviour. The EW of \ion{Si}{xiv} and \ion{S}{xvi} show a similar pattern to \ion{Fe}{xxv} and \ion{Fe}{xxvi} emission lines, decreasing as the egress takes place, reaching a constant value once out-of-eclipse. The intensity of \ion{Mg}{i} and \ion{Si}{xiv} shows a maximum at phase $\phi=0.5$. The EW of \ion{Si}{xi} and \ion{Si}{xiii} shows a suppression before phase 0.2, coinciding with interval 2 of the egress lightcurve, the increase in the color ratio (Fig. \ref{lcurve}) and with the first rise of $N_{\rm H,1}$ (Fig. \ref{disk}).

\begin{figure}
\centering
\subfigure{\includegraphics[trim={6cm 1.6cm 6cm 1.7cm},width=0.9\columnwidth]{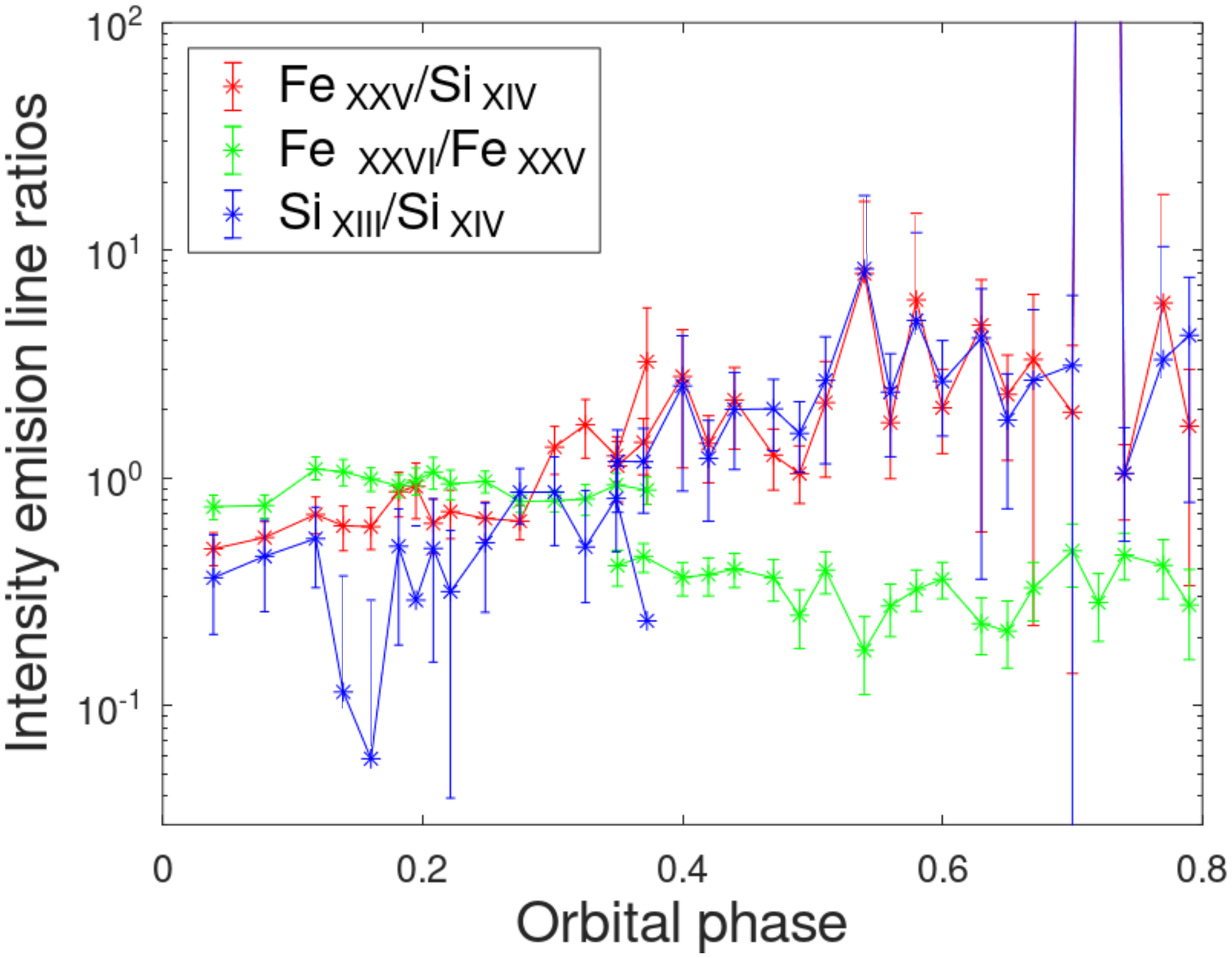}}
\caption{Evolution of the intensity line ratios \ion{Fe}{xxvi}/\ion{Fe}{xxv} (green), \ion{Fe}{xxv}/\ion{Si}{xiv} (red) and \ion{Si}{xiii}/\ion{Si}{xiv} (blue).}
\label{ratios}
\end{figure}

Line intensity ratios can be used to measure the ionization state of the emitting plasma. For that purpose we use the ionization parameter, $\xi$, calculated by \citet[their Fig. 8, upper panel]{8153132}. Our derived parameters are presented in Table \ref{intensity_ratios} and Fig. \ref{ratios}. The \ion{Fe}{xxvi}/\ion{Fe}{xxv} ratio points to a highly ionized plasma with $\log\xi\sim 3.3-3.6$ while \ion{Si}{xiii}/\ion{Si}{xiv} points to $\log\xi\sim 2.1-2.4$, consistent with the values found by \citet{8153132}. This suggest that plasma in different ionization states, probably produced at different sites in the system, are contributing to the observed spectra.

\begin{table}
\caption {Line intensity ratios. The values from this work are averages excluding the most extreme values for each of observation.}
\centering
\begin{adjustbox}{max width=\columnwidth}
\begin{tabular}{cccccc}
\hline\hline
Ratio & \cite{8153132} & This work && This work \\
 & Ratio range & Egress & $\log\xi $& Out-of-eclipse& $\log\xi $ \\
\midrule

\ion{Fe}{xxvi}/\ion{Fe}{xxv}	&	$1-0.5$   &  $\sim$  0.9 & $\sim$ 3.6	&$\sim$ 0.4 &  $\sim$ 3.3\\
\ion{Fe}{xxv}/\ion{Si}{xiv}	&	$1-6.4$   &$\sim$	0.84 & $\sim$2.2  &$\sim$	2.9& $\sim$  3.4	\\
\ion{Si}{xiii}/\ion{Si}{xiv}	&	$0-0.67$  &$\sim$	 0.49 & $\sim$ 2.4 &$\sim$	2.8 & $\sim$ 2.1	\\

\hline
\end{tabular}
\end{adjustbox}
\label{intensity_ratios}
\end{table}

%%%%%%%%%%%%%%%%%%%%%%%%%%%%%%%%%%%%%%%%%%%%%%%%%%%%%
\section{Discussion}

\subsection{\textit{XMM-Newton} lightcurve analysis.} 
\label{disc:sec:timing}

\textit{The egress lightcurve.} The shape of the Cen X-3 egress lightcurve changes in the long term. It appears to be sharp during high states while it shows a more progressive recovery during low states \citep{2008MNRAS.387..439R, 2010RAA....10.1127D}. During the egress (low state) \textit{XMM-Newton} observation, the hard band lightcurve ($3-10$ keV) shows three plateaus separated by two rises, all of them of different duration (Fig. \ref{lcurve}). Could this be caused by decreasing absorption of X-rays in the atmosphere of the companion, as the NS emerges? 

In order to test this suggestion, we computed the opacity of stellar wind for X-rays for typical parameters of an O6.5III star using the state-of-the-art NLTE stellar atmosphere model \textit{PoWR} \citep{Hainich_2019} code\footnote{\texttt{http://www.astro.physik.uni-potsdam.de/$\sim$wrh/PoWR/powrgrid1.php}}. Adopting a wind mass-loss rate $10^{-7}\,M_\odot$\,yr$^{-1}$ and solar metallicity, the stellar wind is transparent to $3-10$ keV X-rays already at 0.03$R_{*}$ above the stellar surface. Thus, the hard band lightcurve should show a steep recovery. This is not compatible with the progressive brightening, particularly during intervals 2 and 4, which takes $\Delta\phi_{\rm orb}\sim$ 0.2. Furthermore, as will be shown in Section \ref{disc:sec:spectra}, although the absorption column decreases overall during egress, it shows enhancements, the most important of which coincides with interval 2 (also visible in the CR, Fig.\ref{lcurve}). 

Another possible explanation is the emergence of an emitting extended structure whose size can be estimated from the corresponding duration $t$ as $v_{\rm o} t \cos{\phi}$ where $v_{\rm o}=436$ km s$^{-1}$ is the orbital velocity of the NS with respect to the system barycentre. The first, more pronounced rise (interval 2 in Fig. \ref{lcurve}) has a duration of $t=12$ ks, which corresponds to a size $l\simeq 0.62 R_{*}$. This is compatible with the Roche lobe of the NS orbiting in a binary system with an orbital separation $a$: $R_{\rm L}(M_{\rm NS}) \simeq 0.46 a (M_{\rm NS}/M_{\rm opt})^{1/3}$. For the Cen X-3 parameters (Table 1) this is $R_{\rm L}(M_{\rm NS})\sim 0.3 R_* \sim l/2$. During this interval of egress, the contributions of the black body and powerlaw components grow. The lightcurve remains constant (interval 3) once the whole structure has emerged out of the eclipse. An obvious candidate for this structure is an accretion disk around the NS. The presence of a disk is required during high states, above the critical luminosity $L_{\rm X}\sim 1 \times 10^{37}$ erg s$^{-1}$ \citep{1979A&A....73...90B}. The shape of the lightcurve suggests that it can also be present at lower luminosity.  

\begin{figure}
\centering
\subfigure{\includegraphics[trim={0cm 4cm 0cm 2cm},width=1\columnwidth]{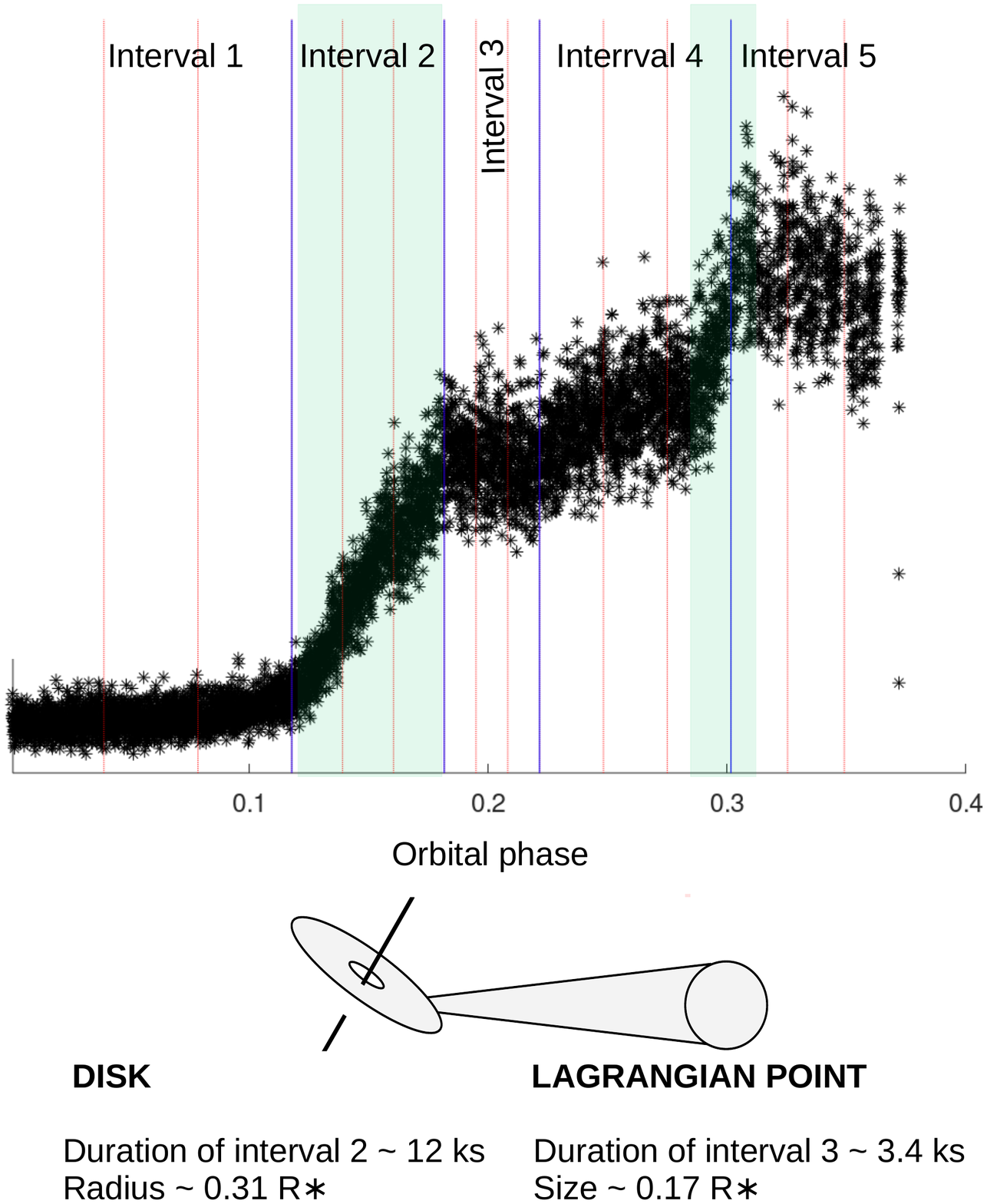}}
\subfigure{\includegraphics[trim={1cm 1cm 3cm 0cm},width=1\columnwidth]{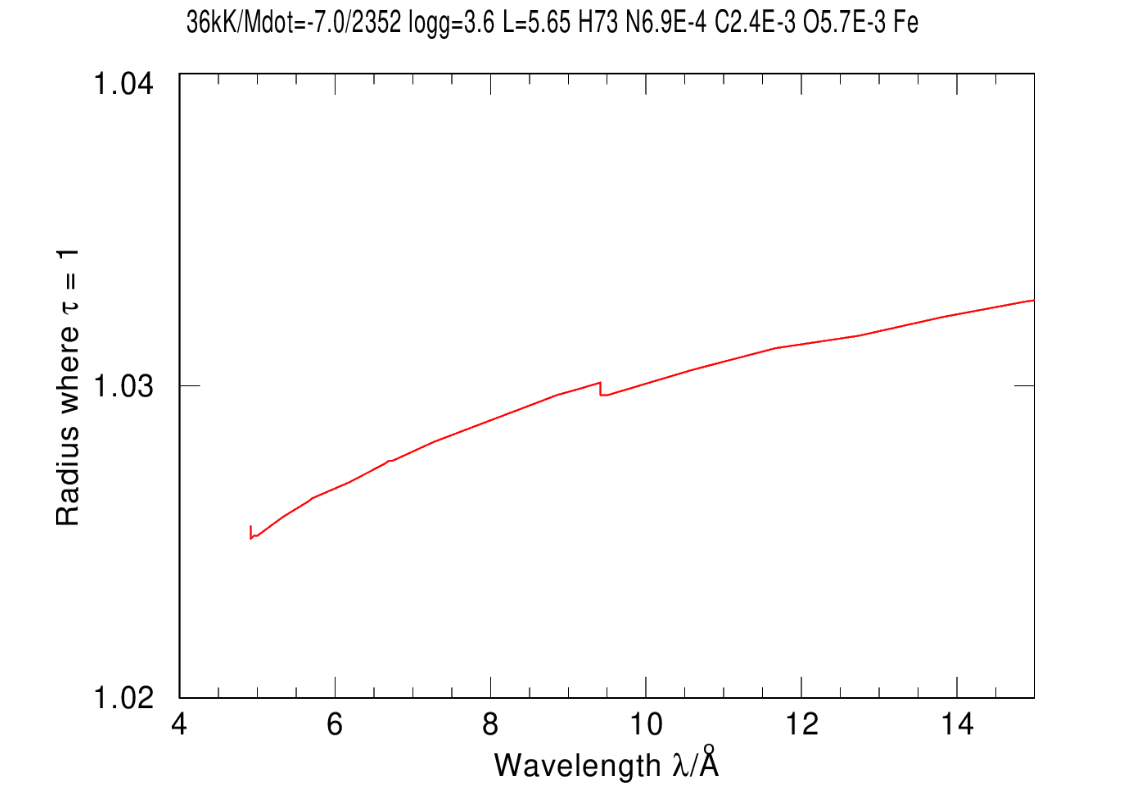}}
\caption { Top panel:
Egress lightcurve and sketch of the accretion disk and hot spot formed during the Roche lobe overflow. Interval 2 is compatible with the disk size. Interval 4, which shows a slight positive increase in the count rate, is produced by the emerging accretion stream. The transition between intervals 4 and 5 shows a short and sharp rise, compatible with a hot spot formed at the Lagrangian point (L1). Bottom panel: radius of a sphere where the optical depth is $\tau=1$ for X-rays, in the $3-10$ keV band, for an O6.5III star atmosphere calculated with {\it PoWR}. }
\label{sketch}
\end{figure}

After that, the lightcurve shows further progressive brightening (interval 4), ended by a sudden moderate rise. These features are, likely, due to other emerging structures. A possible candidate is the accretion stream. The progressive rise could be caused by the impact point of the stream with the disk, or the 'hot line' that appears in the region of the disk-stream interaction, as seen in 3D hydro simulations in \cite{2017MNRAS.467.2934L}. The brightening at the end of interval 4 could be due to the emergence of the Lagrangian point. The brightening from the vicinity of the inner Lagrangian point can be due to X-ray reflection from the optical star atmosphere (Fig. \ref{sketch}). During interval 5, the third plateau occurs when all major emitting sites are visible. This scenario will help us to explain (section \ref{disc:sec:spectra}) the peculiar behaviour of the Fe\,K$\alpha$ line described above.

The low energy lightcurve ($0.2-3$ keV), instead, does not show the structured shape that can be seen in the high energy lightcurve (Fig. \ref{lcurve}, second panel). It is heavily affected by absorption, which will soften any drastic change due to the emerging structures. Rather, it is probably reflecting the general decrease in absorption column as the NS moves towards the observer inside the donor's stellar wind. This is also consistent with the general decrease of $N_{\rm H,1}$ seen in Fig. \ref{disk}. However, in this case, $N_{\rm H,1}$ should rise again after phase 0.5 but this is not observed. 

\textit{Nature of dips}. Two mechanisms can cause the pronounced flux decrease: either an increased absorption or a decrease in mass accretion rate. The first mechanism would require an optically thick structure eclipsing the X-ray source. \citet{2011ApJ...737...79N}, reported the presence of dips during the {\it Suzaku} observation and attributed them to eclipse/obscuration of the X-ray source by clumps of dense matter along the line of sight. However, the dips observed by {\it XMM-Newton} seem to be of different nature. There is no change in the color ratio that could be due to absorption by intervening material. Indeed, the CR should increase during dips as the $0.2-3$ keV band is more suppressed by absorption, contrary to what is observed. Likewise, we should observe an enhancement in the EW of the Fe\,K$\alpha$ line as is regularly seen in other HMXBs \citep[i.e.][]{2010ApJ...715..947T, 2015ApJ...810..102T}, but this is not seen here either.  

 In order to clarify the origin of the dips, we performed a separated phase-resolved spectral analysis for both dips and out-of-dip data (Tables \ref{appendix:C} and \ref{appendix:D}). In Table \ref{dips_spec} we collect phase-averaged model parameters along with their standard deviations. There is no evidence of higher absorption withing the dip duration or, in fact, any other strong differences (including the iron lines; see Tables \ref{appendix:G} and \ref{appendix:H}) except for the flux decrease. 

\begin{table*}
 \caption{Mean values of the spectral-model parameters of the dips and out-of-dip regions. For Fe lines,  $\sigma$ values were set to 1$\times$10$^{-3}$ keV. $L_{39}$ is the luminosity in units of $10^{39}$ erg s$^{-1}$ and $D_{10}$ corresponds to the distance in units of 10 kpc.}
  \centering
    \begin{tabular}{ccc}
\hline\hline

Spectral parameter	&	Dip	&	Out-of-dip	\\
\midrule
				
$N_{\rm H,1}$ ($\times10^{22}$cm$^{-2}$)	&	$3.6\pm0.3$ 		&	$3.7\pm0.3$	\\
$N_{\rm H,2}$	($\times10^{22}$cm$^{-2}$)&	$0.91\pm0.05$		&	$0.93\pm0.06$	\\
$C$	&	$0.80\pm0.03$	&	$0.80\pm0.02$	\\
$\Gamma$	&	$2.27\pm0.21$	&	$2.33\pm0.19$	\\
$K_{\rm po}$ (ph keV$^{-1}$ cm$^{-2}$ s$^{-1}$)	&	$1.4\pm0.3$	&	$1.7\pm0.4$	\\
$kT_{\rm bb}$	(keV)&	$3.03\pm0.21$ 	&	$2.90\pm0.14$\\
$K_{\rm bb}$ ($\times 10^{-2}$ $L_{39}$ $D_{10}^{-2}$)	&	$0.05\pm0.01$ 	&	$0.06\pm0.01$	\\
Flux  ($\times 10^{-10}$ erg cm$^{-2}$ s$^{-1}$)	&	$4.5\pm0.4$	&	$5.0\pm0.4$	\\
CR & $1.43\pm 0.22$ & $1.55 \pm 0.02 $\\
Fe\,K$\alpha$ Centroid (keV) & $6.41\pm0.01$ & $6.41\pm0.01$  \\
 Norm 	($\times10^{-3}$ ph cm$^{-2}$ s$^{-1}$) &  $2.20\pm0.24$  &  $2.39\pm0.19$  \\ 
\ion{Fe}{xxv} Centroid (keV)& $6.67\pm0.01$ & $6.67\pm0.01$\\
 Norm 	($\times10^{-3}$ ph cm$^{-2}$ s$^{-1}$)&  $1.8\pm0.6$ &  $2.0\pm0.5$  \\ 
\ion{Fe}{xxvi} Centroid (keV) &$6.98\pm0.01$  & $6.98\pm0.01$\\
 Norm 	($\times10^{-3}$ ph cm$^{-2}$ s$^{-1}$) & $0.7\pm0.3$ &   $0.7\pm0.3$ \\ 
\hline
    \end{tabular}
  \label{dips_spec}
\end{table*}

Thus, a decrease in the accretion rate is the likely cause for the dips. This could be due to inhomogeneities (rarefactions) in the accretion stream itself.  However, such inhomogeneities would be time spread due to diffusion in the accretion disk. The expected time-scale of such a variability is, at least, of the order of the orbital period. Therefore, we believe that, most likely, the observed dips are due to inevitable instabilities occurring at the inner edge of the disk interacting with the NS magnetosphere (see e.g. the discussion in \citeauthor{2008A&A...480L..21P} \citeyear{2008A&A...480L..21P}). The characteristic time for these instabilities scales with the diffusion time at the inner disk radius $R_{\rm in}$, which is $t_{\rm d}\sim t_{\rm K}(R_{\rm in})\times (h/R_{\rm in})^{-2}\alpha^{-1}$, where $t_{\rm K}$ is the Keplerian time, $h$ is the disk thickness and $\alpha$ is the Shakura-Sunyaev turbulence parameter \citep{1973A&A....24..337S}. Taking the characteristic values for Cen X-3, $t_{\rm K}(R_{\rm in})\approx P_{\rm o}\sim 5$~s, $h/R_{\rm in}\sim 0.1$ and $\alpha \sim 0.3$ gives $1000-1500$ s, which is similar to the observed dip duration. On the other hand, this time is much longer than the duration of the dips observed in direct wind accretors, which are attributed to stellar wind clumps and interclump rarefactions \citep[i.e. $\sim 70\pm 30$ s and $\sim 530\pm 110$ s respectively, for the case of 4U0114+65,][]{2017A&A...606A.145S}. 

\subsection{\textit{XMM-Newton} spectral analysis.} 
\label{disc:sec:spectra}

The spectral fitting outlined in Section \ref{res:sec:spec} uses a phenomenological model composed by a blackbody and a power-law continuum, similar to \citet{Aftab_2019}.
The continuum is modified at low energies by a partial covering absorption (Fig. \ref{disk}). The column $N_{\rm H,2}$ remains constant throughout the two observations at $\simeq 1\times 10^{22}$ cm$^{-2}$. This is compatible with the ISM absorption towards the optical companion ($1.6\pm 0.3\times 10^{22}$ cm$^{-2}$) using $E(B-V)$ from Table \ref{parameters} \citep{2015ApJ...809...66V}. In turn, $N_{\rm H,1}$ (sum of local and ISM), shows a general progressive decrease during the egress before reaching a constant value out-of-eclipse (at $\sim 4\times 10^{22}$ cm$^{-2}$). However, during egress, $N_{\rm H,1}$ shows two important enhancements between orbital phases $\phi = 0.1-0.25$. The first one coincides with interval 2. The absorption increases as the source gets brighter. As explained in Section \ref{disc:sec:timing} this can not be due to the propagation of X-rays through the atmosphere of the donor since it is transparent in the $3-10$ keV band, already at $0.03R_{*}$ above the stellar surface. Thus, there must be local absorbing material corotating with the NS. The covering fraction $C$, a proxy for the donor's stellar wind clumping, varies from 0.76 to 0.9. It is similar to other HMXBs with supergiant donors, where it usually attains values $C>0.85$. This is compatible with the X-ray source being deeply embedded into the stellar wind of the giant star. 

The powerlaw flux is $5-18$ times higher than the black body one (Fig. \ref{disk}) during the low state (egress) and $\sim 1-2$ times higher during the high state (out-of-eclipse). Therefore, it is clear that, although the powerlaw component dominates the Cen X-3 emission at all times, the blackbody contribution is enhanced during high state. The powerlaw photon index $\Gamma$ decreases during the egress observation (Fig. \ref{disk}, lower left). The source becomes progressively harder. However, in the high state (out-of-eclipse) the spectral shape remains constant. $\Gamma$ is also higher so that the source changes from a low-hard to a high-soft state in the long term.

The blackbody temperature is high ($kT\sim 3$ keV) during the high state (out-of-eclipse ). For such high temperature, the displayed luminosity  ($L_{\rm X}\approx 1.8\times 10^{37}$ erg s$^{-1}$), requires a soft seed photon source radius ($A_{\rm spot}=\pi r_{\rm spot}^{2}$) of the order of $r_{\rm spot}\sim 2$ km, compatible with a hot spot on the NS surface. On the other hand, during the low state (egress), when the $L_{\rm X}$ is 10 times lower, the temperature is much lower ($kT\sim 0.4$ keV) and the size of the black-body  emitting area is found to be  much larger, from 10 to 20 km. When the radius of the emitting region exceeds the NS radius, the emitting area should be considered as a sphere ($A_{\rm sphere}=4\pi r_{\rm sphere}^{2}$) instead of a circle. For equal emitting areas $r_\mathrm{sphere}$ = $r_\mathrm{spot}/2$ . Thus, during the low state, the radius would be in the range $\sim 5-10$ km, the upper end being compatible with the NS radius. The much larger fraction of emitting area during the low state also explains the smaller NS pulsed fraction (Fig. \ref{ecpulse}).

\textit{The iron lines.} One of the primary tools for studying the regions of high density in stellar winds is the fluorescence Fe\,K$\alpha$ line from near-neutral Fe \citep{2010ApJ...715..947T, 2015A&A...576A.108G}. Fluorescence arises when matter is illuminated by X-rays. When an iron atom absorbs photons with enough energy to remove an electron from its K-shell ($E>7.2$ keV), the vacancy can be occupied by another electron from an outer shell. If the electron comes from the L-shell, the transition produces Fe\,K$\alpha$ emission. When the ionization state of iron is higher than \ion{Fe}{xix}, the fluorescence yield starts to decrease with the ionization state \citep{2004ApJS..155..675K}. Therefore, Fe\,K$\alpha$ is a footprint of almost neutral Fe (typically less ionized than Fe \textsc{iv}). On the other hand, recombination lines of \ion{Fe}{xxv} and \ion{Fe}{xxvi} unveil the presence of a very hot gas.

\citet{2005ApJ...634L.161I} hypothesizes that the Fe\,K$\alpha$ line should be formed in the outer region of the disk in Cen X-3. \cite{2002xrb..confE..20K}, using \textit{RXTE}, observed a time delay in the variability of the Fe\,K$\alpha$ band, of 0.39 $\pm$ 0.10 ms, and concluded that these photons were reprocessed away from the NS and are, probably, part of the accreting matter. We searched for delays in the time variability of the iron Fe\,K$\alpha$ with respect to \ion{Fe}{xxv} and \ion{Fe}{xxvi} lines using the out-of-eclipse observation. The data was acquired in timing mode and we used a resolution of 0.05 ms. We did not find any delay. The egress observation, in turn, when the lines are much stronger, lacks the required resolution.  \citet{tugay2009xmmnewton}, using the same data set analysed here, conclude that the Fe\,K$\alpha$ line forms in an accretion disk around a NS while highly ionized iron lines form in the outer regions of the binary system. Also, \cite{2012BASI...40..503N} concluded the Fe\,K$\alpha$ line in Cen X-3 is formed by the fluorescence of cold and dense material close to the NS, while highly ionized species are produced in a region far from the NS (in the highly photo-ionized wind of the donor star or in the accretion disk corona). 

As described above, the EW of the highly ionized species (\ion{Fe}{xxv} He like and \ion{Fe}{xxvi} H-like Ly$\alpha$) decrease during the egress. This is expected as the continuum increases dramatically. In contrast, the low ionized Fe\,K$\alpha$ fluorescence line increases during the egress. In line with \citet{Aftab_2019}, our hypothesis is that this line is mostly formed in the channeling stellar wind to the accretion disk, so that it rises as the accretion stream emerges from eclipse. 

In this situation, the intensity of the emission line should show a maximum at the orbital phase $\phi = 0.25$ and a minimum at $\phi = 0.5$, corresponding to maximum and minimum stream projected areas. An intensity peak at $\phi = 0.25$ is indeed clearly detected (see Fig. \ref{felines}, lower panel) but the minimum is not. Unfortunately, the first lightcurve does not continue up to $\phi = 0.5$. Instead, this orbital phase is covered by the second observation when the source was 10 times brighter. A dedicated observation will be needed to further investigate this scenario. 

The intensity of all three Fe lines grow during the egress. \ion{Fe}{xxv} and \ion{Fe}{xxvi} also show the intensity peak at $\phi\sim 0.25$ described before. This means that a significant fraction of their emission must also originate close to the NS and/or along the accretion stream. In the high state, the intensity of all three lines follow a similar pattern. However, \ion{Fe}{xxv} experiences a sudden drop at $\phi\sim 0.7$. This is intriguing. Since the spectral parameters of the X-ray source remain rather constant during the high state, this drop must be related to geometrical factors (i.e. \ion{Fe}{xxv} emitting region subtending a smaller area close to NS quadrature). The intensity of \ion{Mg}{i} and \ion{Si}{xiv}, in turn, shows a maximum at phase 0.5. An interesting possibility is that they are formed at the irradiated face of the donor.

%%%%%%%%%%%%%%%%%%%%%%%%%%%%%%%%%%%%%%%%%%%%%%
\section{Summary and conclusions}

\begin{enumerate}

\item We describe the X-ray spectra of Cen X-3 by a phenomenological model consisting of a black body plus a power law. The use of alternative thermal components (i.e. \texttt{diskbb}) results in poorer statistics. Thermal comptonisation (i.e. \texttt{comptt}) or hybrid thermal plus bulk comptonisation models (i.e. \texttt{compmag}) do not describe well the data either. Purely thermal models (e.g., the disk emission) fit the data well but the parameters are difficult to explain or unphysical. 

\item The source exhibits high-soft $\Leftrightarrow$ low-hard transitions. During high state, the black body component increases in the overall emission budget although the power law dominates the whole spectrum at all fluxes. The black-body emitting area has an equivalent radius of the order of 2 km, compatible with a hot spot on the NS surface. During low states, the size of the black-body emitting area is found to be much larger, from 5 to 10 km. Concurrently, the pulsed fraction is also smaller during low states. 

\item The absorption column to the X-ray source has two components. On one hand, $N_{\rm H,2}$ $\sim 0.5-1\times 10^{22}$ cm$^{-2}$, is compatible with the ISM, as deduced from optical observations of the donor. On the other hand, $N_{\rm H,1}$, corresponds to the local plus ISM absorption. It decreases as the egress progresses, from $N_{\rm H,1}$ $\sim 7-8\times 10^{22}$ cm$^{-2}$ to $N_{\rm H,1}$ $\sim 4\times 10^{22}$ cm$^{-2}$ out-of-eclipse. However, it displays two enhancements, coinciding with the two flux rises (intervals 2 and 4, Fig. \ref{lcurve}). Thus, the local material is absorbing the X-ray source as it emerges from the eclipse, probably corotating with it. 

\item The spectra show emission lines from a photoionised plasma. The most prominent lines correspond to Fe. The equivalent width of highly ionized species (\ion{Fe}{xxv} He like, and \ion{Fe}{xxvi} H-like Ly$\alpha$) decrease during the egress as the continuum rises. Their intensities increase during the egress, thereby demonstrating that a significant fraction must originate relatively close to the NS. In turn, the equivalent width of the neutral Fe increases. It must originate in dense and cold structures emerging during the egress. We suggest that these structures are located along the accretion stream. In this scenario, the line intensities must show maximum and minimum at orbital phases 0.25 and 0.5 respectively, corresponding to maximum and minimum stream projected areas. While the first one is seen, the second is not, although the source was in different states.

\item The low state egress $3-10$ keV ligtcurve is highly structured showing several intervals. This structure can not be explained by the propagation of X-rays through the stellar wind of the donor but, rather, is due to the appearance of several extended emitting regions. The first, most prominent rise, corresponds to a structure whose size is compatible with the Roche lobe size of Cen X-3, $R_{\rm L}\simeq 0.3R_{*}$. The second rise signals the egress of an emitting region with a size $\simeq 0.17R_{*}$. Possible candidates for these structures could be the accretion disk and reflection from the the optical star atmosphere close to the inner Lagrangian point, respectively.  

\item The out-of-eclipse lightcurve (high state) shows prominent dips. These dips are not caused by absorption of the intervening material. The separate spectral analysis for both dips and out-of-dip data found no significant differences in the absorption columns or any other parameter except the flux, which is reduced by a $\sim40\%$. Instead, they can be caused by a sporadic decrease in the accretion rate, most likely, due to instabilities at the inner edge of the disk interacting with the NS magnetosphere. The characteristic time for these instabilities scales with the diffusion time at the inner disk radius which, for Cen X-3, turns out to be $t_{\rm d}\sim 1.0-1.5\times 10^{3}$ s, close to the observed dip duration. This is much longer than the dips seen in direct wind accretors, attributed to wind clumps (tens of seconds) and interclump rarefactions (hundreds of seconds).  

\end{enumerate}

\section*{Acknowledgements}
This research has been supported by the project ESP2017-85691-P
LMO acknowledges Deutsches Zentrum für Luft und Raumfahrt (DLR) grant FKZ
50 OR 1508 and partial support by the Russian Government Program of
Competitive Growth of Kazan Federal University. J.J.R.R. acknowledges financial support from the Spanish Ministry of
Education, Culture and Sport fellowship PRX17/00114, and also thanks
all the staff from SRON for their collaboration and hospitality there.
KP acknowledges support from RFBR grant 18-502-12025. 
We acknowledge the constructive criticism of the referee whose comments improved the content of the paper.

%%%%%%%%%%%%%%%%%%%%%%%%%%%%%%%%%%%%%%%%%%%%%%%%%%
\section*{Data Availability}
The data analysed in this study can be found in the {\it XMM-Newton} archive under the observation identification numbers 0111010101 and 0400550201, for the eclipse egress observation and the out-of-eclipse observation respectively.



\bibliographystyle{mnras}
\bibliography{bib.bib} % if your bibtex file is called example.bib

%\begin{thebibliography}{99}
%\end{thebibliography}

%%%%%%%%%%%%%%%%%%%%%%%%%%%%%%%%%%%%%%%%%%%%%%%%%%

%%%%%%%%%%%%%%%%% APPENDICES %%%%%%%%%%%%%%%%%%%%%

\newpage

\onecolumn

\appendix

\section{Phase resolved spectral parameters.}

\begin{table*}
\centering
\caption{Phase-resolved spectral parameters for the egress observation. The first column refers to intervals in Fig. \ref{lcurve}. The number of degrees of freedom for all spectra is 234. $L_{39}$ is the luminosity in units of $10^{39}$ erg s$^{-1}$ and $D_{10}$ corresponds to the distance in units of 10 kpc.}
\begin{adjustbox}{max width=\textwidth}
\begin{tabular}{cccccccccc}
\hline\hline
	Spectra  & Orb. phase 	&	$ \chi ^2 $	&	 $N_{\rm H,1}$	&	$N_{\rm H,2}$	&	 $C$ 	&	 $\Gamma$ 	&	$K_{\rm po}$	&	 $kT_{\rm bb}$	& $K_{\rm bb}$	\\
	&	&			&	($\times10^{22}$ cm$^{-2}$) 	&	($\times10^{22}$ cm$^{-2}$) 	&		&	&	 ($\times 10^{-3}$ ph keV$^{-1}$ cm $^{-2}$s$^{-1}$ )		&	(keV)	&	  ($\times 10^{-4}$ $L_{39}$ $D_{10}^{-2}$ )	\\
	\midrule
1.1 	&	 0.03 	&	 1.01 	&	 $7.4_{-0.7}^{+0.7}$ 	&	 $0.70_{-0.04}^{+0.05}$ 	&	 $0.13_{-0.02}^{+0.02}$ 	&	 $1.00_{-0.11}^{+0.10}$ 	&	 $3.6_{-0.7}^{+0.7}$ 	&	 $0.43_{-0.05}^{+0.05}$ 	&	 $3.4_{-0.6}^{+0.8}$\\\\
1.2 	&	 0.08 	&	 1.12 	&	 $6.7_{-0.5}^{+0.5}$ 	&	 $0.78_{-0.05}^{+0.05}$ 	&	 $0.11_{-0.10}^{+0.02}$ 	&	 $1.09_{-0.08}^{+0.06}$ 	&	 $4.4_{-0.6}^{+0.6}$ 	&	 $0.38_{-0.01}^{+0.04}$ 	&	 $4.1_{-1.1}^{+0.4}$\\\\
1.3 	&	 0.12 	&	 1.40 	&	 $6.0_{-0.5}^{+0.5}$ 	&	 $0.63_{-0.04}^{+0.05}$ 	&	 $0.13_{-0.02}^{+0.02}$ 	&	 $0.97_{-0.08}^{+0.08}$ 	&	 $4.1_{-0.6}^{+0.6}$ 	&	 $0.42_{-0.04}^{+0.05}$ 	&	 $2.8_{-0.6}^{+0.8}$\\\\
2.1 	&	 0.14 	&	 0.90 	&	 $6.5_{-0.6}^{+0.6}$ 	&	 $0.73_{-0.06}^{+0.06}$ 	&	 $0.10_{-0.10}^{+0.02}$ 	&	 $0.61_{-0.06}^{+0.06}$ 	&	 $4.1_{-0.5}^{+0.5}$ 	&	 $0.38_{-0.03}^{+0.04}$ 	&	 $5.7_{-1.4}^{+0.7}$\\\\
2.2 	&	 0.16 	&	 1.07 	&	 $8.3_{-0.6}^{+0.7}$ 	&	 $1.02_{-0.04}^{+0.05}$ 	&	 $0.10_{-0.10}^{+0.02}$ 	&	 $0.54_{-0.05}^{+0.05}$ 	&	 $6.8_{-0.6}^{+0.7}$ 	&	 $0.41_{-0.03}^{+0.04}$ 	&	 $8.7_{-1.5}^{+0.8}$\\\\
2.3 	&	 0.18 	&	 1.25 	&	 $7.3_{-0.6}^{+0.7}$ 	&	 $0.86_{-0.05}^{+0.06}$ 	&	 $0.10_{-0.10}^{+0.01}$ 	&	 $0.54_{-0.04}^{+0.04}$ 	&	 $9.0_{-0.7}^{+0.7}$ 	&	 $0.38_{-0.04}^{+0.04}$ 	&	 $8.0_{-1.5}^{+1.0}$\\\\
3.1 	&	 0.19 	&	 1.14 	&	 $7.3_{-0.6}^{+0.6}$ 	&	 $0.80_{-0.06}^{+0.07}$ 	&	 $0.10_{-0.10}^{+0.02}$ 	&	 $0.42_{-0.06}^{+0.06}$ 	&	 $8.3_{-0.8}^{+0.9}$ 	&	 $0.45_{-0.04}^{+0.04}$ 	&	 $9.2_{-1.9}^{+1.4}$\\\\
3.2 	&	 0.21 	&	 0.98 	&	 $6.9_{-0.6}^{+0.6}$ 	&	 $0.83_{-0.06}^{+0.06}$ 	&	 $0.10_{-0.10}^{+0.03}$ 	&	 $0.53_{-0.06}^{+0.05}$ 	&	 $10.1_{-1.0}^{+1.0}$ 	&	 $0.42_{-0.04}^{+0.06}$ 	&	 $10.0_{-2.4}^{+1.4}$\\\\
3.3 	&	 0.22 	&	 1.11 	&	 $6.0_{-0.4}^{+0.4}$ 	&	 $0.78_{-0.05}^{+0.05}$ 	&	 $0.10_{-0.10}^{+0.02}$ 	&	 $0.58_{-0.05}^{+0.05}$ 	&	 $10.4_{-1.0}^{+1.0}$ 	&	 $0.43_{-0.03}^{+0.04}$ 	&	 $8.5_{-1.8}^{+1.4}$\\\\
4.1 	&	 0.25 	&	 1.19 	&	 $6.6_{-0.3}^{+0.3}$ 	&	 $0.80_{-0.04}^{+0.04}$ 	&	 $0.10_{-0.10}^{+0.02}$ 	&	 $0.59_{-0.04}^{+0.03}$ 	&	 $12.0_{-1.0}^{+1.0}$ 	&	 $0.43_{-0.02}^{+0.03}$ 	&	 $13.9_{-2.4}^{+1.3}$\\\\
4.2 	&	 0.27 	&	 0.95 	&	 $6.9_{-0.3}^{+0.3}$ 	&	 $0.80_{-0.03}^{+0.03}$ 	&	 $0.10_{-0.10}^{+0.01}$ 	&	 $0.61_{-0.04}^{+0.02}$ 	&	 $13.2_{-1.0}^{+0.5}$ 	&	 $0.42_{-0.01}^{+0.02}$ 	&	 $20_{-3}^{+1}$\\\\
4.3 	&	 0.30 	&	 1.06 	&	 $5.2_{-0.3}^{+0.3}$ 	&	 $0.63_{-0.04}^{+0.05}$ 	&	 $0.10_{-0.10}^{+0.01}$ 	&	 $0.19_{-0.04}^{+0.04}$ 	&	 $6.9_{-0.6}^{+0.6}$ 	&	 $0.43_{-0.01}^{+0.02}$ 	&	 $15.3_{-2.4}^{+1.7}$\\\\
5.1 	&	 0.33 	&	 0.79 	&	 $4.9_{-0.5}^{+0.6}$ 	&	 $0.67_{-0.07}^{+0.08}$ 	&	 $0.13_{-0.03}^{+0.04}$ 	&	 $-0.22_{-0.08}^{+0.07}$ 	&	 $6.2_{-0.8}^{+0.9}$ 	&	 $0.45_{-0.04}^{+0.04}$ 	&	 $22_{-4}^{+6}$\\\\
5.2 	&	 0.35 	&	 1.13 	&	 $5.4_{-0.4}^{+0.4}$ 	&	 $0.63_{-0.04}^{+0.04}$ 	&	 $0.14_{-0.03}^{+0.03}$ 	&	 $0.05_{-0.06}^{+0.05}$ 	&	 $8.2_{-0.9}^{+0.9}$ 	&	 $0.44_{-0.03}^{+0.04}$ 	&	 $20_{-4}^{+6}$\\\\
5.3 	&	 0.37 	&	 0.96 	&	 $5.1_{-0.4}^{+0.5}$ 	&	 $0.59_{-0.06}^{+0.06}$ 	&	 $0.13_{-0.02}^{+0.02}$ 	&	 $0.4_{-0.6}^{+0.5}$ 	&	 $10.4_{-1.0}^{+1.1}$ 	&	 $0.41_{-0.03}^{+0.04}$ 	&	 $11.8_{-3}^{+4}$\\

\hline
\end{tabular}
\end{adjustbox}
\label{appendix:A}
\end{table*}

\newpage
%\section{phase-resolved spectroscopy for the out-of-eclipse observation. }

\begin{table*}
\centering
\caption{Phase-resolved spectral parameters for the out-of-eclipse observation.  The number of degrees of freedom for all spectra is 1873. $L_{39}$ is the luminosity in units of $10^{39}$ erg s$^{-1}$ and $D_{10}$ corresponds to the distance in units of 10 kpc.}
\begin{adjustbox}{max width=\textwidth}
\begin{tabular}{cccccccccc}
\hline\hline 
	Spectra  & 	Orb. phase 	&	$ \chi ^2 $	&	 $N_{\rm H,1}$	&	$N_{\rm H,2}$ &	 $C$ 	&	$ \Gamma$ 	&	$K_{\rm po}$	&	 $kT_{\rm bb}$	& $K_{\rm bb}$	\\
	&	&			&	($\times10^{22}$ cm$^{-2}$) 	&	($\times10^{22}$ cm$^{-2}$) 	&		&	&	 (ph keV$^{-1}$ cm $^{-2}$s$^{-1}$ )		&	(keV)	&	  ($\times 10^{-2}$ $L_{39}$ $D_{10}^{-2}$ )	\\
\midrule
1 	&	 0.35 	&	 1.14 	&	 $3.78_{-0.20}^{+0.09}$ 	&	 $0.92_{-0.02}^{+0.01}$ 	&	 $0.78_{-0.01}^{+0.01}$ 	&	 $2.21_{-0.16}^{+0.13}$ 	&	 $1.7_{-0.3}^{+0.2}$ 	&	 $3.00_{-0.05}^{+0.06}$ 	&	 $6.2_{-0.4}^{+0.2}$ \\\\
2 	&	 0.37 	&	 1.08 	&	 $3.6_{-0.5}^{+0.3}$ 	&	 $0.90_{-0.06}^{+0.05}$ 	&	 $0.79_{-0.03}^{+0.02}$ 	&	 $2.19_{-0.22}^{+0.16}$ 	&	 $1.7_{-0.4}^{+0.4}$ 	&	 $3.01_{-0.06}^{+0.11}$ 	&	 $6.2_{-0.8}^{+0.4}$ \\\\
3 	&	 0.40 	&	 1.13 	&	 $3.7_{-0.5}^{+0.3}$ 	&	 $0.91_{-0.06}^{+0.04}$ 	&	 $0.80_{-0.03}^{+0.02}$ 	&	 $2.2_{-0.3}^{+0.2}$ 	&	 $1.7_{-0.5}^{+0.4}$ 	&	 $3.03_{-0.06}^{+0.12}$ 	&	 $6.3_{-0.8}^{+0.4}$ \\\\
4 	&	 0.42 	&	 1.07 	&	 $3.9_{-0.3}^{+0.2}$ 	&	 $0.95_{-0.05}^{+0.03}$ 	&	 $0.81_{-0.02}^{+0.02}$ 	&	 $2.41_{-0.16}^{+0.15}$ 	&	 $2.0_{-0.4}^{+0.4}$ 	&	 $2.91_{-0.06}^{+0.07}$ 	&	 $6.5_{-0.2}^{+0.3}$ \\\\

5 	&	 0.44 	&	 1.02 	&	 $3.3_{-0.3}^{+0.3}$ 	&	 $0.85_{-0.05}^{+0.05}$ 	&	 $0.76_{-0.01}^{+0.02}$ 	&	 $1.95_{-0.10}^{+0.24}$ 	&	 $1.18_{-0.23}^{+0.24}$ 	&	 $3.16_{-0.09}^{+0.09}$ 	&	 $5.2_{-0.6}^{+0.8}$ \\\\
6 	&	 0.47 	&	 1.13 	&	 $3.89_{-0.11}^{+0.08}$ 	&	 $0.93_{-0.04}^{+0.02}$ 	&	 $0.78_{-0.01}^{+0.02}$ 	&	 $2.23_{-0.07}^{+0.02}$ 	&	 $1.74_{-0.24}^{+0.19}$ 	&	 $3.00_{-0.07}^{+0.04}$ 	&	 $6.2_{-0.4}^{+0.2}$ \\\\
7 	&	 0.49 	&	 1.14 	&	 $3.8_{-0.3}^{+0.2}$ 	&	 $0.93_{-0.05}^{+0.03}$ 	&	 $0.80_{-0.03}^{+0.01}$ 	&	 $2.30_{-0.22}^{+0.07}$ 	&	 $1.8_{-0.4}^{+0.3}$ 	&	 $2.98_{-0.07}^{+0.10}$ 	&	 $6.1_{-0.6}^{+0.3}$ \\\\
8 	&	 0.51 	&	 1.17 	&	 $3.4_{-0.4}^{+0.2}$ 	&	 $0.85_{-0.07}^{+0.03}$ 	&	 $0.78_{-0.03}^{+0.01}$ 	&	 $2.04_{-0.24}^{+0.15}$ 	&	 $1.3_{-0.3}^{+0.2}$ 	&	 $3.02_{-0.06}^{+0.10}$ 	&	 $5.4_{-1.0}^{+0.4}$ \\\\
9 	&	 0.54 	&	 1.14 	&	 $3.48_{-0.16}^{+0.24}$ 	&	 $0.89_{-0.02}^{+0.04}$ 	&	 $0.80_{-0.02}^{+0.01}$ 	&	 $2.22_{-0.16}^{+0.15}$ 	&	 $1.5_{-0.2}^{+0.3}$ 	&	 $2.94_{-0.03}^{+0.07}$ 	&	 $5.8_{-0.4}^{+0.3}$ \\\\
10 	&	 0.56 	&	 1.06 	&	 $3.88_{-0.20}^{+0.16}$ 	&	 $0.94_{-0.03}^{+0.01}$ 	&	 $0.79_{-0.03}^{+0.01}$ 	&	 $2.29_{-0.11}^{+0.05}$ 	&	 $1.75_{-0.22}^{+0.13}$ 	&	 $3.00_{-0.04}^{+0.06}$ 	&	 $6.0_{-0.4}^{+0.1}$ \\\\
11 	&	 0.58 	&	 1.12 	&	 $3.5_{-0.3}^{+0.2}$ 	&	 $0.90_{-0.06}^{+0.03}$ 	&	 $0.78_{-0.04}^{+0.02}$ 	&	 $2.12_{-0.22}^{+0.09}$ 	&	 $1.4_{-0.3}^{+0.2}$ 	&	 $3.05_{-0.07}^{+0.05}$ 	&	 $5.6_{-0.6}^{+0.3}$ \\\\
12 	&	 0.60 	&	 1.12 	&	 $3.8_{-0.4}^{+0.1}$ 	&	 $0.94_{-0.04}^{+0.02}$ 	&	 $0.78_{-0.02}^{+0.02}$ 	&	 $2.27_{-0.19}^{+0.13}$ 	&	 $1.7_{-0.4}^{+0.3}$ 	&	 $3.08_{-0.09}^{+0.10}$ 	&	 $6.1_{-0.5}^{+0.2}$ \\\\
13 	&	 0.63 	&	 1.07 	&	 $3.56_{-0.09}^{+0.22}$ 	&	 $0.89_{-0.02}^{+0.06}$ 	&	 $0.77_{-0.01}^{+0.02}$ 	&	 $2.10_{-0.12}^{+0.24}$ 	&	 $1.4_{-0.2}^{+0.4}$ 	&	 $3.13_{-0.11}^{+0.08}$ 	&	 $5.5_{-0.4}^{+0.4}$ \\\\
14 	&	 0.65 	&	 1.10 	&	 $3.4_{-0.3}^{+0.3}$ 	&	 $0.86_{-0.03}^{+0.04}$ 	&	 $0.77_{-0.02}^{+0.02}$ 	&	 $2.08_{-0.17}^{+0.21}$ 	&	 $1.2_{-0.2}^{+0.4}$ 	&	 $3.06_{-0.08}^{+0.08}$ 	&	 $5.1_{-0.3}^{+0.5}$ \\\\
15 	&	 0.67 	&	 1.15 	&	 $3.5_{-0.3}^{+0.2}$ 	&	 $0.89_{-0.03}^{+0.04}$ 	&	 $0.80_{-0.02}^{+0.02}$ 	&	 $2.22_{-0.20}^{+0.16}$ 	&	 $1.3_{-0.3}^{+0.3}$ 	&	 $2.93_{-0.07}^{+0.09}$ 	&	 $5.0_{-0.3}^{+0.3}$ \\\\
16 	&	 0.70 	&	 1.09 	&	 $3.9_{-0.5}^{+0.2}$ 	&	 $0.96_{-0.07}^{+0.04}$ 	&	 $0.82_{-0.02}^{+0.02}$ 	&	 $2.51_{-0.24}^{+0.15}$ 	&	 $1.6_{-0.4}^{+0.4}$ 	&	 $2.69_{-0.04}^{+0.07}$ 	&	 $5.0_{-0.4}^{+0.2}$ \\\\
17 	&	 0.72 	&	 1.07 	&	 $3.6_{-0.3}^{+0.3}$ 	&	 $0.9_{-0.05}^{+0.04}$ 	&	 $0.81_{-0.02}^{+0.02}$ 	&	 $2.30_{-0.18}^{+0.20}$ 	&	 $1.4_{-0.3}^{+0.4}$ 	&	 $2.79_{-0.05}^{+0.06}$ 	&	 $5.1_{-0.2}^{+0.3}$ \\\\
18 	&	 0.74 	&	 1.09 	&	 $3.8_{-0.2}^{+0.3}$ 	&	 $0.94_{-0.04}^{+0.05}$ 	&	 $0.80_{-0.02}^{+0.02}$ 	&	 $2.36_{-0.16}^{+0.21}$ 	&	 $1.6_{-0.3}^{+0.4}$ 	&	 $2.85_{-0.05}^{+0.05}$ 	&	 $5.3_{-0.4}^{+0.3}$ \\\\
19 	&	 0.77 	&	 1.15 	&	 $3.4_{-0.3}^{+0.5}$ 	&	 $0.88_{-0.04}^{+0.06}$ 	&	 $0.79_{-0.01}^{+0.02}$ 	&	 $2.20_{-0.18}^{+0.21}$ 	&	 $1.2_{-0.3}^{+0.3}$ 	&	 $2.81_{-0.05}^{+0.03}$ 	&	 $4.7_{-0.7}^{+0.4}$ \\\\
20	&	0.79	&	1.14	&	$3.88_{-0.24}^{+0.23}$	&	$0.94_{-0.04}^{+0.04}$	&	$0.81_{-0.02}^{+0.02}$	&	$2.45_{-0.17}^{+0.17}$	&	$1.7_{-0.3}^{+0.4}$	&	$2.71_{-0.05}^{+0.04}$	&	$5.5_{-0.3}^{+0.2}$	\\

\hline
\end{tabular}
\end{adjustbox}
\label{appendix:B}
\end{table*}

\newpage
%\section{phase-resolved spectra for the dips in the out-of-eclipse A. }
\begin{table*}
\centering
\caption{Phase-resolved spectral parameters for the dips. The number of degrees of freedom is 166 for dips 23 and 26, 167 for dips 9, 14, 18, 25 and 27, 168 for dips 1, 10, 13, 16, 20 and 21 and 169 for the rest. $L_{39}$ is the luminosity in units of $10^{39}$ erg s$^{-1}$ and $D_{10}$ corresponds to the distance in units of 10 kpc.}
\begin{adjustbox}{max width=\textwidth}
\begin{tabular}{cccccccccc}
\hline\hline 
	Spectra  & 	Orb. phase 	&	$ \chi ^2 $	&	 $N_{\rm H,1}$	&	$N_{\rm H,2}$   &	 $C$ 	&	 $\Gamma$ 	&	 $K_{\rm po}$	&	 $kT_{\rm bb}$	& $K_{\rm bb}$	\\
	&	&			&	($\times10^{22}$ cm$^{-2}$) 	&	($\times10^{22}$ cm$^{-2}$) 	&		&	&	 (ph keV$^{-1}$ cm $^{-2}$s$^{-1}$ )		&	(keV)	&	  ($\times 10^{-2}$ $L_{39}$ $D_{10}^{-2}$ )	\\
	\midrule
1	&	0.35	&	1.24	&	$3.39_{-0.24}^{+0.24}$	&	$0.91_{-0.04}^{+0.04}$	&	$0.72_{-0.04}^{+0.04}$	&	$1.93_{-0.18}^{+0.20}$	&	$1.1_{-0.2}^{+0.3}$	&	$3.5_{-0.3}^{+0.4}$	&	$5.1_{-0.8}^{+0.6}$	\\\\
2	&	0.37	&	1.17	&	$3.2_{-0.24}^{+0.24}$	&	$0.86_{-0.05}^{+0.05}$	&	$0.77_{-0.02}^{+0.02}$	&	$2.05_{-0.18}^{+0.19}$	&	$1.2_{-0.2}^{+0.3}$	&	$3.03_{-0.12}^{+0.15}$	&	$5.2_{-0.6}^{+0.5}$	\\\\
3	&	0.39	&	1.12	&	$4.08_{-0.17}^{+0.18}$	&	$0.99_{-0.03}^{+0.03}$	&	$0.80_{-0.02}^{+0.02}$	&	$2.43_{-0.13}^{+0.14}$	&	$2.0_{-0.3}^{+0.4}$	&	$3.16_{-0.13}^{+0.15}$	&	$6.2_{-0.3}^{+0.3}$	\\\\
4	&	0.41	&	1.49	&	$2.98_{-0.21}^{+0.21}$	&	$0.81_{-0.04}^{+0.04}$	&	$0.78_{-0.01}^{+0.01}$	&	$1.89_{-0.14}^{+0.14}$	&	$1.01_{-0.15}^{+0.18}$	&	$3.29_{-0.14}^{+0.18}$	&	$5.1_{-0.6}^{+0.5}$	\\\\
5	&	0.43	&	1.27	&	$3.90_{-0.16}^{+0.17}$	&	$0.97_{-0.02}^{+0.03}$	&	$0.77_{-0.02}^{+0.02}$	&	$2.23_{-0.13}^{+0.14}$	&	$1.5_{-0.2}^{+0.3}$	&	$3.21_{-0.14}^{+0.18}$	&	$5.7_{-0.4}^{+0.3}$	\\\\
6	&	0.44	&	1.45	&	$3.69_{-0.14}^{+0.14}$	&	$0.92_{-0.02}^{+0.02}$	&	$0.79_{-0.02}^{+0.02}$	&	$2.21_{-0.11}^{+0.12}$	&	$1.50_{-0.17}^{+0.21}$	&	$3.10_{-0.12}^{+0.14}$	&	$5.7_{-0.3}^{+0.3}$	\\\\
7	&	0.46	&	1.04	&	$3.61_{-0.17}^{+0.18}$	&	$0.91_{-0.02}^{+0.03}$	&	$0.75_{-0.03}^{+0.03}$	&	$2.09_{-0.13}^{+0.14}$	&	$1.32_{-0.18}^{+0.23}$	&	$3.21_{-0.16}^{+0.21}$	&	$5.7_{-0.5}^{+0.4}$	\\\\
8	&	0.46	&	1.03	&	$3.73_{-0.19}^{+0.20}$	&	$0.92_{-0.03}^{+0.04}$	&	$0.80_{-0.02}^{+0.02}$	&	$2.27_{-0.15}^{+0.16}$	&	$1.7_{-0.3}^{+0.3}$	&	$3.10_{-0.13}^{+0.16}$	&	$6.2_{-0.4}^{+0.4}$	\\\\
9	&	0.5	&	1.01	&	$3.4_{-0.3}^{+0.3}$	&	$0.88_{-0.05}^{+0.05}$	&	$0.78_{-0.04}^{+0.03}$	&	$2.13_{-0.23}^{+0.24}$	&	$1.3_{-0.2}^{+0.4}$	&	$3.2_{-0.2}^{+0.4}$	&	$5.5_{-0.8}^{+0.6}$	\\\\
10	&	0.52	&	1.14	&	$2.65_{-0.19}^{+0.22}$	&	$0.74_{-0.04}^{+0.04}$	&	$0.80_{-0.02}^{+0.02}$	&	$1.82_{-0.14}^{+0.15}$	&	$0.87_{-0.12}^{+0.16}$	&	$3.08_{-0.16}^{+0.21}$	&	$4.5_{-0.6}^{+0.6}$	\\\\
11	&	0.53	&	1.28	&	$3.63_{-0.23}^{+0.24}$	&	$0.94_{-0.04}^{+0.04}$	&	$0.82_{-0.02}^{+0.02}$	&	$2.39_{-0.16}^{+0.16}$	&	$1.7_{-0.2}^{+0.3}$	&	$2.99_{-0.12}^{+0.15}$	&	$5.6_{-0.4}^{+0.3}$	\\\\
12	&	0.56	&	1.15	&	$3.70_{-0.16}^{+0.17}$	&	$0.93_{-0.03}^{+0.03}$	&	$0.81_{-0.02}^{+0.02}$	&	$2.35_{-0.13}^{+0.14}$	&	$1.6_{-0.2}^{+0.3}$	&	$3.01_{-0.13}^{+0.15}$	&	$5.5_{-0.3}^{+0.3}$	\\\\
13	&	0.58	&	1.12	&	$3.3_{-0.3}^{+0.3}$	&	$0.91_{-0.06}^{+0.06}$	&	$0.80_{-0.03}^{+0.02}$	&	$2.31_{-0.20}^{+0.22}$	&	$1.5_{-0.2}^{+0.4}$	&	$2.89_{-0.13}^{+0.16}$	&	$5.5_{-0.6}^{+0.5}$	\\\\
14	&	0.59	&	0.86	&	$3.4_{-0.3}^{+0.3}$	&	$0.93_{-0.05}^{+0.06}$	&	$0.80_{-0.04}^{+0.03}$	&	$2.42_{-0.23}^{+0.24}$	&	$1.6_{-0.3}^{+0.5}$	&	$2.84_{-0.15}^{+0.19}$	&	$5.8_{-0.5}^{+0.5}$	\\\\
15	&	0.6	&	1.23	&	$3.35_{-0.22}^{+0.24}$	&	$0.88_{-0.04}^{+0.04}$	&	$0.79_{-0.02}^{+0.02}$	&	$2.12_{-0.15}^{+0.17}$	&	$1.3_{-0.2}^{+0.3}$	&	$3.16_{-0.16}^{+0.20}$	&	$5.3_{-0.5}^{+0.5}$	\\\\
16	&	0.62	&	1.01	&	$3.9_{-0.3}^{+0.3}$	&	$0.97_{-0.05}^{+0.05}$	&	$0.78_{-0.03}^{+0.03}$	&	$2.30_{-0.21}^{+0.23}$	&	$1.65_{-0.32}^{+0.45}$	&	$3.2_{-0.2}^{+0.3}$	&	$5.8_{-0.6}^{+0.5}$	\\\\
17	&	0.63	&	1.46	&	$4.15_{-0.15}^{+0.15}$	&	$1.00_{-0.02}^{+0.03}$	&	$0.79_{-0.02}^{+0.02}$	&	$2.39_{-0.12}^{+0.13}$	&	$1.8_{-0.2}^{+0.3}$	&	$3.21_{-0.12}^{+0.15}$	&	$6.01_{-0.24}^{+0.24}$	\\\\
18	&	0.65	&	0.93	&	$3.2_{-0.3}^{+0.3}$	&	$0.85_{-0.05}^{+0.05}$	&	$0.75_{-0.03}^{+0.03}$	&	$1.91_{-0.19}^{+0.22}$	&	$0.95_{-0.17}^{+0.24}$	&	$3.1_{-0.2}^{+0.3}$	&	$4.4_{-0.9}^{+0.7}$	\\\\
19	&	0.66	&	1.09	&	$3.55_{-0.20}^{+0.20}$	&	$0.88_{-0.03}^{+0.04}$	&	$0.77_{-0.02}^{+0.02}$	&	$2.14_{-0.15}^{+0.16}$	&	$1.16_{-0.18}^{+0.24}$	&	$3.24_{-0.14}^{+0.20}$	&	$4.9_{-0.4}^{+0.4}$	\\\\
20	&	0.68	&	1.1	&	$3.78_{-0.20}^{+0.21}$	&	$0.97_{-0.03}^{+0.03}$	&	$0.74_{-0.03}^{+0.03}$	&	$2.17_{-0.16}^{+0.18}$	&	$1.2_{-0.2}^{+0.3}$	&	$3.33_{-0.18}^{+0.24}$	&	$4.9_{-0.4}^{+0.4}$	\\\\
21	&	0.68	&	1.07	&	$3.3_{-0.3}^{+0.3}$	&	$0.88_{-0.05}^{+0.05}$	&	$0.83_{-0.02}^{+0.02}$	&	$2.36_{-0.19}^{+0.21}$	&	$1.2_{-0.2}^{+0.3}$	&	$2.88_{-0.11}^{+0.14}$	&	$4.6_{-0.3}^{+0.3}$	\\\\
22	&	0.69	&	1.22	&	$3.3_{-0.3}^{+0.4}$	&	$0.87_{-0.07}^{+0.07}$	&	$0.85_{-0.01}^{+0.01}$	&	$2.52_{-0.23}^{+0.24}$	&	$1.3_{-0.3}^{+0.4}$	&	$2.59_{-0.06}^{+0.06}$	&	$4.6_{-0.3}^{+0.2}$	\\\\
23	&	0.7	&	0.91	&	$3.7_{-0.4}^{+0.4}$	&	$0.93_{-0.05}^{+0.06}$	&	$0.81_{-0.03}^{+0.03}$	&	$2.3_{-0.3}^{+0.3}$	&	$1.3_{-0.3}^{+0.6}$	&	$2.89_{-0.16}^{+0.22}$	&	$4.5_{-0.5}^{+0.4}$	\\\\
24	&	0.71	&	1.35	&	$3.59_{-0.24}^{+0.24}$	&	$0.91_{-0.04}^{+0.04}$	&	$0.79_{-0.03}^{+0.02}$	&	$2.25_{-0.19}^{+0.19}$	&	$1.2_{-0.2}^{+0.3}$	&	$2.95_{-0.11}^{+0.13}$	&	$4.7_{-0.4}^{+0.3}$	\\\\
25	&	0.72	&	0.77	&	$4.1_{-0.3}^{+0.3}$	&	$0.96_{-0.06}^{+0.06}$	&	$0.85_{-0.02}^{+0.02}$	&	$2.67_{-0.23}^{+0.24}$	&	$1.9_{-0.4}^{+0.6}$	&	$2.71_{-0.11}^{+0.13}$	&	$5.3_{-0.3}^{+0.3}$	\\\\
26	&	0.73	&	0.87	&	$3.6_{-0.2}^{+0.3}$	&	$0.92_{-0.04}^{+0.04}$	&	$0.81_{-0.03}^{+0.03}$	&	$2.39_{-0.19}^{+0.21}$	&	$1.5_{-0.3}^{+0.4}$	&	$2.98_{-0.16}^{+0.21}$	&	$5.3_{-0.4}^{+0.4}$	\\\\
27	&	0.73	&	1.13	&	$3.4_{-0.4}^{+0.4}$	&	$0.90_{-0.07}^{+0.07}$	&	$0.82_{-0.03}^{+0.03}$	&	$2.4_{-0.2}^{+0.3}$	&	$1.4_{-0.4}^{+0.6}$	&	$2.81_{-0.12}^{+0.14}$	&	$5.1_{-0.4}^{+0.4}$	\\\\
28	&	0.75	&	1.2	&	$3.89_{-0.17}^{+0.18}$	&	$0.96_{-0.03}^{+0.03}$	&	$0.81_{-0.02}^{+0.02}$	&	$2.43_{-0.13}^{+0.14}$	&	$1.5_{-0.2}^{+0.3}$	&	$2.88_{-0.09}^{+0.11}$	&	$4.85_{-0.23}^{+0.21}$	\\\\
29	&	0.76	&	1.26	&	$3.63_{-0.21}^{+0.23}$	&	$0.93_{-0.04}^{+0.04}$	&	$0.81_{-0.02}^{+0.02}$	&	$2.40_{-0.15}^{+0.16}$	&	$1.3_{-0.2}^{+0.3}$	&	$2.96_{-0.10}^{+0.12}$	&	$4.68_{-0.24}^{+0.22}$	\\\\
30	&	0.78	&	1.41	&	$3.89_{-0.24}^{+0.24}$	&	$0.98_{-0.04}^{+0.05}$	&	$0.84_{-0.02}^{+0.02}$	&	$2.64_{-0.19}^{+0.19}$	&	$1.8_{-0.4}^{+0.5}$	&	$2.69_{-0.07}^{+0.08}$	&	$5.07_{-0.23}^{+0.19}$	\\\\
31	&	0.79	&	1.31	&	$3.57_{-0.18}^{+0.19}$	&	$0.93_{-0.03}^{+0.04}$	&	$0.82_{-0.02}^{+0.02}$	&	$2.44_{-0.14}^{+0.15}$	&	$1.40_{-0.19}^{+0.24}$	&	$2.73_{-0.07}^{+0.07}$	&	$5.07_{-0.23}^{+0.21}$	\\
\hline
\end{tabular}
\end{adjustbox}
\label{appendix:C}
\end{table*}

\newpage
%\section{phase-resolved spectroscopy for the out-of-dip periods of the out-of-eclipse observation. }

\begin{table*}
\centering
\caption{Phase-resolved spectral parameters for the out-of-dip periods of the out-of-eclipse observation. The number of degrees of freedom is 167 for bins 20 and 23, 168 for bins 5, 6, 9, 10, 11, 17 and 21, and 169 for the rest. $L_{39}$ is the luminosity in units of $10^{39}$ erg s$^{-1}$ and $D_{10}$ corresponds to the distance in units of 10 kpc.}  
\begin{adjustbox}{max width=\textwidth}
\begin{tabular}{cccccccccc}
\hline\hline
	Spectra  & 	Orb. phase  &	$ \chi ^2 $	&	 $N_{\rm H,1}$	  &	$N_{\rm H,2}$	  &	 $C$ 	&	 $\Gamma$ 	&	 $K_{\rm po}$	&	 $kT_{\rm bb}$	& $K_{\rm bb}$	\\
	&	&			&	($\times10^{22}$ cm$^{-2}$) 	&	($\times10^{22}$ cm$^{-2}$) 	&		&	&	 (ph keV$^{-1}$ cm $^{-2}$s$^{-1}$ )		&	(keV)	&	  ($\times 10^{-2}$ $L_{39}$ $D_{10}^{-2}$ )	\\	
\midrule
1	&	0.36	&	1.95	&	$3.75_{-0.12}^{+0.12}$	&	$0.93_{-0.02}^{+0.02}$	&	$0.80_{-0.01}^{+0.01}$		&	$2.27_{-0.10}^{+0.10}$	&	$1.81_{-0.19}^{+0.22}$		&	$2.95_{-0.06}^{+0.07}$	&	$6.5_{-0.3}^{+0.2}$	\\\\
2	&	0.38	&	1.48	&	$3.78_{-0.15}^{+0.16}$	&	$0.93_{-0.03}^{+0.03}$	&	$0.80_{-0.02}^{+0.02}$		&	$2.31_{-0.12}^{+0.13}$	&	$1.9_{-0.2}^{+0.3}$	    	&	$2.94_{-0.08}^{+0.09}$	&	$6.6_{-0.3}^{+0.3}$	\\\\
3	&	0.42	&	1.27	&	$4.10_{-0.12}^{+0.13}$	&	$0.99_{-0.02}^{+0.02}$	&	$0.82_{-0.01}^{+0.01}$		&	$2.56_{-0.11}^{+0.11}$	&	$2.4_{-0.3}^{+0.4}$	    	&	$2.88_{-0.06}^{+0.07}$	&	$6.84_{-0.20}^{+0.19}$	\\\\
4	&	0.43	&	1.18	&	$3.85_{-0.12}^{+0.13}$	&	$0.96_{-0.02}^{+0.02}$	&	$0.80_{-0.02}^{+0.02}$		&	$2.37_{-0.10}^{+0.11}$	&	$1.9_{-0.2}^{+0.3}$ 		&	$2.89_{-0.07}^{+0.08}$	&	$6.51_{-0.24}^{+0.23}$	\\\\
5	&	0.45	&	1.26	&	$3.79_{-0.16}^{+0.17}$	&	$0.96_{-0.03}^{+0.03}$	&	$0.81_{-0.02}^{+0.02}$		&	$2.38_{-0.12}^{+0.13}$	&	$1.9_{-0.2}^{+0.3}$	    	&	$2.96_{-0.09}^{+0.11}$	&	$6.3_{-0.3}^{+0.3}$	\\\\
6	&	0.46	&	1.22	&	$3.48_{-0.22}^{+0.24}$	&	$0.87_{-0.04}^{+0.04}$	&	$0.79_{-0.03}^{+0.02}$		&	$2.02_{-0.16}^{+0.18}$	&	$1.4_{-0.2}^{+0.3}$	    	&	$3.2_{-0.2}^{+0.3}$	    &	$5.9_{-0.7}^{+0.7}$	\\\\
7	&	0.5	&	1.87	&	$3.48_{-0.13}^{+0.13}$  	&	$0.88_{-0.02}^{+0.02}$	&	$0.80_{-0.01}^{+0.01}$		&	$2.18_{-0.09}^{+0.10}$	&	$1.45_{-0.15}^{+0.17}$		&	$3.05_{-0.07}^{+0.08}$	&	$5.83_{-0.24}^{+0.24}$	\\\\
8	&	0.52	&	1.19	&	$3.61_{-0.15}^{+0.15}$	&	$0.90_{-0.02}^{+0.03}$	&	$0.78_{-0.02}^{+0.02}$		&	$2.14_{-0.11}^{+0.12}$	&	$1.44_{-0.17}^{+0.21}$		&	$2.98_{-0.09}^{+0.10}$	&	$5.7_{-0.4}^{+0.3}$	    \\\\
9	&	0.57	&	1.92	&	$3.82_{-0.09}^{+0.09}$	&	$0.95_{-0.01}^{+0.02}$	&	$0.79_{-0.01}^{+0.01}$		&	$2.28_{-0.07}^{+0.07}$	&	$1.68_{-0.13}^{+0.15}$		&	$3.02_{-0.06}^{+0.06}$	&	$5.99_{-0.17}^{+0.16}$	\\\\
10	&	0.58	&	1.25	&	$4.0_{-0.2}^{+0.3}$	    &	$1.00_{-0.04}^{+0.05}$	&	$0.81_{-0.03}^{+0.02}$		&	$2.44_{-0.19}^{+0.21}$	&	$2.1_{-0.4}^{+0.6}$	    	&	$2.96_{-0.12}^{+0.15}$	&	$6.5_{-0.4}^{+0.4}$	\\\\
11	&	0.59	&	0.96	&	$3.7_{-0.2}^{+0.3}$	    &	$0.94_{-0.04}^{+0.05}$	&	$0.81_{-0.02}^{+0.02}$		&	$2.35_{-0.19}^{+0.20}$	&	$1.8_{-0.4}^{+0.5}$ 		&	$2.92_{-0.12}^{+0.14}$	&	$6.4_{-0.5}^{+0.4}$	\\\\
12	&	0.6	&	1.75	&	$3.47_{-0.13}^{+0.13}$	    &	$0.91_{-0.02}^{+0.02}$	&	$0.78_{-0.01}^{+0.01}$		&	$2.14_{-0.09}^{+0.10}$	&	$1.38_{-0.14}^{+0.16}$		&	$3.11_{-0.08}^{+0.09}$	&	$5.8_{-0.3}^{+0.3}$	\\\\
13	&	0.62	&	1.02	&	$3.91_{-0.16}^{+0.16}$	&	$0.96_{-0.03}^{+0.03}$	&	$0.79_{-0.02}^{+0.02}$		&	$2.32_{-0.13}^{+0.13}$	&	$1.8_{-0.2}^{+0.3}$		&	$3.09_{-0.10}^{+0.12}$	&	$6.3_{-0.3}^{+0.3}$	\\\\
14	&	0.64	&	1.44	&	$3.39_{-0.19}^{+0.19}$	&	$0.89_{-0.03}^{+0.03}$	&	$0.79_{-0.02}^{+0.02}$		&	$2.17_{-0.14}^{+0.14}$	&	$1.38_{-0.20}^{+0.24}$		&	$2.96_{-0.07}^{+0.08}$	&	$5.6_{-0.4}^{+0.3}$	\\\\
15	&	0.65	&	1.36	&	$3.61_{-0.16}^{+0.16}$	&	$0.91_{-0.03}^{+0.03}$	&	$0.77_{-0.02}^{+0.02}$		&	$2.19_{-0.13}^{+0.13}$	&	$1.38_{-0.18}^{+0.22}$		&	$3.03_{-0.09}^{+0.11}$	&	$5.5_{-0.4}^{+0.3}$	\\\\
16	&	0.66	&	1.24	&	$3.52_{-0.20}^{+0.22}$	&	$0.88_{-0.04}^{+0.04}$	&	$0.80_{-0.02}^{+0.02}$		&	$2.23_{-0.16}^{+0.16}$	&	$1.3_{-0.2}^{+0.3}$	    	&	$2.94_{-0.10}^{+0.12}$	&	$5.2_{-0.4}^{+0.3}$	\\\\
17	&	0.67	&	1.01	&	$2.8_{-0.3}^{+0.4}$	    &	$0.76_{-0.07}^{+0.07}$	&	$0.80_{-0.02}^{+0.02}$		&	$1.89_{-0.22}^{+0.24}$	&	$0.9_{-0.2}^{+0.3}$	    	&	$2.96_{-0.13}^{+0.16}$	&	$4.5_{-0.8}^{+0.7}$	\\\\
18	&	0.68	&	0.8	&	$4.02_{-0.21}^{+0.23}$  	&	$0.98_{-0.04}^{+0.04}$	&	$0.82_{-0.02}^{+0.02}$		&	$2.52_{-0.17}^{+0.19}$	&	$1.8_{-0.3}^{+0.5}$	    	&	$2.78_{-0.08}^{+0.09}$	&	$5.5_{-0.3}^{+0.2}$	\\\\
19	&	0.69	&	1.12	&	$3.9_{-0.3}^{+0.3}$	    &	$0.96_{-0.05}^{+0.05}$	&	$0.79_{-0.04}^{+0.03}$		&	$2.35_{-0.24}^{+0.22}$	&	$1.4_{-0.4}^{+0.4}$		    &	$2.82_{-0.09}^{+0.12}$	&	$5.2_{-0.5}^{+0.3}$	\\\\
20	&	0.7	&	0.92	&	$3.61_{-0.22}^{+0.24}$  	&	$0.91_{-0.03}^{+0.04}$	&	$0.80_{-0.03}^{+0.03}$		&	$2.27_{-0.18}^{+0.20}$	&	$1.3_{-0.2}^{+0.3}$	    	&	$2.77_{-0.12}^{+0.15}$	&	$4.8_{-0.4}^{+0.4}$	\\\\
21	&	0.71	&	1.32	&	$3.7_{-0.3}^{+0.3}$	    &	$0.95_{-0.05}^{+0.05}$	&	$0.81_{-0.03}^{+0.03}$		&	$2.44_{-0.22}^{+0.23}$	&	$1.6_{-0.3}^{+0.5}$	    	&	$2.73_{-0.09}^{+0.10}$	&	$5.3_{-0.4}^{+0.4}$	\\\\
22	&	0.72	&	1.15	&	$4.08_{-0.17}^{+0.17}$	&	$1.01_{-0.03}^{+0.03}$	&	$0.82_{-0.02}^{+0.02}$		&	$2.60_{-0.13}^{+0.14}$	&	$2.0_{-0.3}^{+0.4}$	    	&	$2.78_{-0.07}^{+0.08}$	&	$5.51_{-0.20}^{+0.19}$	\\\\
23	&	0.72	&	0.93	&	$3.1_{-0.5}^{+0.6}$	    &	$0.84_{-0.14}^{+0.14}$	&	$0.86_{-0.02}^{+0.02}$		&	$2.5_{-0.3}^{+0.4}$	    &	$1.6_{-0.5}^{+1.0}$	    	&	$2.54_{-0.11}^{+0.14}$	&	$5.4_{-0.6}^{+0.5}$	\\\\
24	&	0.73	&	1.56	&	$3.96_{-0.11}^{+0.12}$	&	$0.97_{-0.02}^{+0.02}$	&	$0.81_{-0.01}^{+0.01}$		&	$2.48_{-0.09}^{+0.10}$	&	$1.82_{-0.18}^{+0.22}$		&	$2.84_{-0.05}^{+0.06}$	&	$5.67_{-0.16}^{+0.15}$	\\\\
25	&	0.75	&	1.14	&	$3.91_{-0.19}^{+0.21}$	&	$0.96_{-0.03}^{+0.04}$	&	$0.81_{-0.02}^{+0.02}$		&	$2.44_{-0.16}^{+0.17}$	&	$1.7_{-0.3}^{+0.4}$	    	&	$2.88_{-0.09}^{+0.11}$	&	$5.5_{-0.3}^{+0.3}$	\\\\
26	&	0.76	&	1.34	&	$3.94_{-0.15}^{+0.16}$	&	$0.97_{-0.02}^{+0.02}$	&	$0.79_{-0.02}^{+0.02}$		&	$2.38_{-0.12}^{+0.13}$	&	$1.61_{-0.19}^{+0.24}$		&	$2.77_{-0.08}^{+0.09}$	&	$5.5_{-0.3}^{+0.3}$	\\\\
27	&	0.77	&	1.43	&	$4.33_{-0.14}^{+0.15}$	&	$1.03_{-0.03}^{+0.03}$	&	$0.85_{-0.01}^{+0.01}$		&	$2.81_{-0.12}^{+0.13}$	&	$2.5_{-0.3}^{+0.4}$	    	&	$2.64_{-0.05}^{+0.05}$	&	$5.92_{-0.14}^{+0.14}$	\\
\hline
\end{tabular}
\end{adjustbox}
\label{appendix:D}
\end{table*}

\begin{table*}
\centering
\caption{Fe lines gaussian parameters for the egress.$\sigma$ values compatible with zero were set to 1$\times$10$^{-3}$ keV.}
\begin{adjustbox}{max width=\textwidth}
\begin{tabular}{ccccccccccc}
\hline\hline																					
Spectra	&	Orb. Phase	&	Fe\,K$\alpha$ Centroid	&	Fe\,K$\alpha$ $\sigma$	& Fe\,K$\alpha$ Norm	&	\ion{Fe}{xxv} Centroid	&\ion{Fe}{xxv} $\sigma$	&\ion{Fe}{xxv} Norm	&	\ion{Fe}{xxvi} Centroid	&	\ion{Fe}{xxvi} $\sigma$	&	\ion{Fe}{xxvi} Norm	\\\\
	&		&	keV	&	eV	&	($\times10^{-4}$ ph cm$^{-2}$ s$^{-1}$)	&	keV	&	eV	&	($\times10^{-4}$ ph cm$^{-2}$ s$^{-1}$)	&	keV	&	eV	&	($\times10^{-4}$ ph cm$^{-2}$ s$^{-1}$)	\\\\
		\midrule	
1.1	&	0.01	&	$6.41_{-0.03}^{+0.03}$	&	$9_{-3}^{+9}$	&	$1.04_{-0.22}^{+0.20}$	&	$6.67_{-0.01}^{+0.01}$	&	1	&	$2.24_{-0.16}^{+0.22}$	&	$6.97_{-0.01}^{+0.01}$	&	1	&	$1.68_{-0.17}^{+0.85}$	\\\\
1.2	&	0.04	&	$6.40_{-0.02}^{+0.02}$	&	$5_{-4}^{+4}$	&	$0.90_{-0.20}^{+0.23}$	&	$6.67_{-0.01}^{+0.01}$	&	1	&	$2.55_{-0.20}^{+0.20}$	&	$6.97_{-0.01}^{+0.01}$	&	1	&	$1.87_{-0.17}^{+0.00}$	\\\\
1.3	&	0.08	&	$6.41_{-0.03}^{+0.03}$	&	$10_{-4}^{+8}$	&	$1.31_{-0.24}^{+0.19}$	&	$6.67_{-0.01}^{+0.01}$	&	1	&	$2.45_{-0.17}^{+0.20}$	&	$6.97_{-0.01}^{+0.01}$	&	1	&	$2.46_{-0.19}^{+0.16}$	\\\\
2.1	&	0.11	&	$6.39_{-0.02}^{+0.02}$	&	$10_{-4}^{+10}$	&	$2.8_{-0.5}^{+0.4}$	&	$6.68_{-0.01}^{+0.01}$	&	1	&	$3.4_{-0.3}^{+0.4}$	&	$6.98_{-0.01}^{+0.01}$	&	1	&	$3.4_{-0.3}^{+0.2}$	\\\\
2.2	&	0.13	&	$6.40_{-0.01}^{+0.01}$	&	$7.5_{-2.0}^{+2.1}$	&	$5.3_{-0.6}^{+0.6}$	&	$6.68_{-0.01}^{+0.01}$	&	1	&	$5.5_{-0.4}^{+0.2}$	&	$6.97_{-0.01}^{+0.01}$	&	1	&	$4.8_{-0.4}^{+0.2}$	\\\\
2.3	&	0.15	&	$6.40_{-0.01}^{+0.01}$	&	$5.1_{-1.9}^{+2.2}$	&	$6.2_{-0.6}^{+0.6}$	&	$6.67_{-0.01}^{+0.01}$	&	1	&	$6.5_{-0.5}^{+0.4}$	&	$6.97_{-0.01}^{+0.01}$	&	1	&	$5.1_{-0.5}^{+0.3}$	\\\\
3.1	&	0.16	&	$6.40_{-0.01}^{+0.01}$	&	$7.4_{-2.3}^{+2.4}$	&	$8.8_{-0.9}^{+1.0}$	&	$6.68_{-0.01}^{+0.01}$	&	1	&	$8.2_{-0.8}^{+0.5}$	&	$6.97_{-0.02}^{+0.01}$	&	1	&	$6.1_{-0.7}^{+0.4}$	\\\\
3.2	&	0.17	&	$6.41_{-0.01}^{+0.01}$	&	$6_{-3}^{+3}$	&	$7.6_{-0.9}^{+1.0}$	&	$6.67_{-0.01}^{+0.01}$	&	1	&	$5.9_{-0.8}^{+0.5}$	&	$6.97_{-0.02}^{+0.02}$	&	1	&	$5.4_{-0.6}^{+0.5}$	\\\\
3.3	&	0.19	&	$6.41_{-0.01}^{+0.01}$	&	$7.0_{-1.7}^{+1.8}$	&	$8.5_{-0.9}^{+0.9}$	&	$6.68_{-0.01}^{+0.01}$	&	1	&	$6.1_{-0.7}^{+0.7}$	&	$6.97_{-0.01}^{+0.01}$	&	1	&	$5.3_{-0.6}^{+0.7}$	\\\\
4.1	&	0.21	&	$6.40_{-0.01}^{+0.01}$	&	$7.8_{-1.9}^{+1.9}$	&	$9.3_{-0.6}^{+0.4}$	&	$6.68_{-0.01}^{+0.01}$	&	1	&	$6.0_{-0.5}^{+0.7}$	&	$6.97_{-0.01}^{+0.01}$	&	1	&	$5.7_{-0.4}^{+0.6}$	\\\\
4.2	&	0.24	&	$6.40_{-0.01}^{+0.01}$	&	$5.9_{-1.5}^{+1.4}$	&	$10.1_{-0.6}^{+0.6}$	&	$6.68_{-0.01}^{+0.01}$	&	$5_{-3}^{+2}$	&	$8.1_{-0.6}^{+0.7}$	&	$6.98_{-0.01}^{+0.01}$	&	$5.5_{-2.0}^{+2.0}$	&	$6.4_{-0.5}^{+0.6}$	\\\\
4.3	&	0.27	&	$6.40_{-0.01}^{+0.01}$	&	$4.9_{-1.6}^{+1.2}$	&	$14.6_{-0.8}^{+0.8}$	&	$6.68_{-0.01}^{+0.01}$	&	$4_{-3}^{+2}$	&	$10.7_{-0.7}^{+0.5}$	&	$6.98_{-0.01}^{+0.01}$	&	$5.3_{-2.4}^{+2.1}$	&	$8.5_{-0.7}^{+0.5}$	\\\\
5.1	&	0.29	&	$6.40_{-0.01}^{+0.01}$	&	1	&	$22.8_{-2.1}^{+2.1}$	&	$6.67_{-0.01}^{+0.02}$	&	$7_{-1}^{+5}$	&	$25.5_{-2.4}^{+0.6}$	&	$6.98_{-0.02}^{+0.02}$	&	$7_{-3}^{+3}$	&	$20.6_{-2.4}^{+0.5}$	\\\\
5.2	&	0.32	&	$6.41_{-0.01}^{+0.01}$	&	$6.1_{-1.6}^{+1.3}$	&	$23.3_{-1.7}^{+1.7}$	&	$6.68_{-0.01}^{+0.01}$	&	1	&	$15.5_{-1.3}^{+0.7}$	&	$6.99_{-0.01}^{+0.01}$	&	1	&	$12.7_{-1.2}^{+0.7}$	\\\\
5.3	&	0.34	&	$6.41_{-0.01}^{+0.01}$	&	$6.8_{-1.5}^{+1.3}$	&	$16.5_{-1.0}^{+1.0}$	&	$6.68_{-0.01}^{+0.01}$	&	$0.2_{-0.2}^{+4}$	&	$9.5_{-0.8}^{+2.4}$	&	$6.98_{-0.01}^{+0.01}$	&	$5.2_{-2.3}^{+2.0}$	&	$8.4_{-0.8}^{+2.4}$	\\\\

\bottomrule																
\end{tabular}																
\end{adjustbox}																
\label{appendix:E}																	
\end{table*}

\begin{table*}
\centering
\caption{Fe lines gaussian parameters for the out-of-eclipse observation. All $\sigma$ values were set to 1$\times$10$^{-3}$ keV.}
\begin{adjustbox}{max width=\textwidth}
\begin{tabular}{cccccccc}
\hline\hline
%Out-of-elcipse	&		&		&		&		&		&		&			\\\\
Spectra	&	Orb. Phase	&	Fe\,K$\alpha$ Centroid	& Fe\,K$\alpha$ Norm	&	\ion{Fe}{xxv} Centroid	&	\ion{Fe}{xxv} Norm	&	\ion{Fe}{xxvi} Centroid &	\ion{Fe}{xxvi} Norm		\\
	&		&	(keV)	&	($\times10^{-3}$ ph cm$^{-2}$ s$^{-1}$)	&	(keV)	&($\times10^{-3}$ ph cm$^{-2}$ s$^{-1}$)	&	(keV)	&	($\times10^{-3}$ ph cm$^{-2}$ s$^{-1}$)		\\
		\midrule
1	&	0.35	&	$6.41_{-0.01}^{+0.01}$	&	$2.55_{-0.16}^{+0.18}$	&	$6.68_{-0.01}^{+0.01}$	&	$2.36_{-0.17}^{+0.16}$	&	$6.97_{-0.03}^{+0.02}$	&	$0.97_{-0.15}^{+0.17}$	\\\\
2	&	0.37	&	$6.41_{-0.01}^{+0.01}$	&	$2.69_{-0.16}^{+0.18}$	&	$6.67_{-0.01}^{+0.01}$	&	$2.65_{-0.17}^{+0.17}$	&	$6.97_{-0.02}^{+0.02}$	&	$1.2_{-0.15}^{+0.17}$	\\\\
3	&	0.40	&	$6.41_{-0.01}^{+0.01}$	&	$2.75_{-0.17}^{+0.17}$	&	$6.68_{-0.01}^{+0.01}$	&	$2.76_{-0.16}^{+0.17}$	&	$6.97_{-0.03}^{+0.02}$	&	$1.01_{-0.15}^{+0.16}$	\\\\
4	&	0.42	&	$6.41_{-0.01}^{+0.01}$	&	$2.54_{-0.16}^{+0.16}$	&	$6.68_{-0.01}^{+0.01}$	&	$2.38_{-0.16}^{+0.16}$	&	$6.97_{-0.04}^{+0.03}$	&	$0.90_{-0.16}^{+0.16}$	\\\\
5	&	0.44	&	$6.41_{-0.01}^{+0.01}$	&	$2.55_{-0.15}^{+0.16}$	&	$6.67_{-0.01}^{+0.01}$	&	$2.51_{-0.17}^{+0.16}$	&	$6.97_{-0.04}^{+0.03}$	&	$1.00_{-0.16}^{+0.16}$	\\\\
6	&	0.47	&	$6.41_{-0.01}^{+0.01}$	&	$2.58_{-0.16}^{+0.17}$	&	$6.68_{-0.02}^{+0.01}$	&	$2.27_{-0.17}^{+0.16}$	&	$6.97_{-0.03}^{+0.02}$	&	$0.83_{-0.16}^{+0.16}$	\\\\
7	&	0.49	&	$6.41_{-0.01}^{+0.01}$	&	$2.42_{-0.17}^{+0.16}$	&	$6.66_{-0.02}^{+0.01}$	&	$2.09_{-0.16}^{+0.16}$	&	$6.97_{-0.03}^{+0.05}$	&	$0.52_{-0.15}^{+0.14}$	\\\\
8	&	0.51	&	$6.41_{-0.01}^{+0.01}$	&	$2.53_{-0.16}^{+0.17}$	&	$6.67_{-0.01}^{+0.02}$	&	$2.11_{-0.17}^{+0.17}$	&	$6.97_{-0.03}^{+0.02}$	&	$0.83_{-0.15}^{+0.17}$	\\\\
9	&	0.54	&	$6.41_{-0.01}^{+0.01}$	&	$2.52_{-0.17}^{+0.15}$	&	$6.66_{-0.03}^{+0.02}$	&	$2.25_{-0.15}^{+0.17}$	&	$6.97_{-0.03}^{+0.02}$	&	$0.39_{-0.16}^{+0.14}$	\\\\
10	&	0.56	&	$6.41_{-0.01}^{+0.01}$	&	$2.74_{-0.15}^{+0.17}$	&	$6.68_{-0.01}^{+0.01}$	&	$2.24_{-0.17}^{+0.15}$	&	$6.97_{-0.03}^{+0.03}$	&	$0.61_{-0.15}^{+0.16}$	\\\\
11	&	0.58	&	$6.40_{-0.01}^{+0.01}$	&	$2.83_{-0.17}^{+0.15}$	&	$6.66_{-0.01}^{+0.01}$	&	$2.40_{-0.15}^{+0.17}$	&	$6.97_{-0.04}^{+0.02}$	&	$0.78_{-0.15}^{+0.14}$	\\\\
12	&	0.60	&	$6.40_{-0.01}^{+0.01}$	&	$2.50_{-0.17}^{+0.16}$	&	$6.66_{-0.01}^{+0.01}$	&	$2.44_{-0.16}^{+0.17}$	&	$6.97_{-0.04}^{+0.03}$	&	$0.88_{-0.15}^{+0.15}$	\\\\
13	&	0.63	&	$6.40_{-0.01}^{+0.01}$	&	$2.47_{-0.16}^{+0.15}$	&	$6.66_{-0.02}^{+0.01}$	&	$2.39_{-0.16}^{+0.17}$	&	$6.97_{-0.03}^{+0.04}$	&	$0.55_{-0.16}^{+0.14}$	\\\\
14	&	0.65	&	$6.40_{-0.01}^{+0.01}$	&	$2.40_{-0.17}^{+0.15}$	&	$6.66_{-0.02}^{+0.01}$	&	$2.07_{-0.15}^{+0.17}$	&	$6.97_{-0.03}^{+0.03}$	&	$0.44_{-0.15}^{+0.14}$	\\\\
15	&	0.67	&	$6.40_{-0.01}^{+0.01}$	&	$2.56_{-0.16}^{+0.15}$	&	$6.66_{-0.02}^{+0.02}$	&	$1.59_{-0.15}^{+0.15}$	&	$6.97_{-0.03}^{+0.02}$	&	$0.52_{-0.15}^{+0.14}$	\\\\
16	&	0.70	&	$6.40_{-0.01}^{+0.01}$	&	$2.53_{-0.15}^{+0.14}$	&	$6.66_{-0.02}^{+0.02}$	&	$1.04_{-0.14}^{+0.15}$	&	$6.98_{-0.03}^{+0.02}$	&	$0.50_{-0.14}^{+0.13}$	\\\\
17	&	0.72	&	$6.40_{-0.01}^{+0.01}$	&	$2.53_{-0.16}^{+0.15}$	&	$6.66_{-0.03}^{+0.01}$	&	$1.53_{-0.15}^{+0.15}$	&	$6.97_{-0.06}^{+0.02}$	&	$0.43_{-0.14}^{+0.13}$	\\\\
18	&	0.74	&	$6.40_{-0.01}^{+0.01}$	&	$2.52_{-0.16}^{+0.15}$	&	$6.66_{-0.03}^{+0.02}$	&	$1.51_{-0.14}^{+0.16}$	&	$6.97_{-0.06}^{+0.02}$	&	$0.69_{-0.15}^{+0.14}$	\\\\
19	&	0.77	&	$6.41_{-0.01}^{+0.01}$	&	$2.47_{-0.15}^{+0.15}$	&	$6.68_{-0.02}^{+0.02}$	&	$1.28_{-0.14}^{+0.14}$	&	$6.97_{-0.03}^{+0.03}$	&	$0.53_{-0.14}^{+0.14}$	\\\\
20	&	0.79	&	$6.41_{-0.01}^{+0.01}$	&	$2.40_{-0.15}^{+0.15}$	&	$6.68_{-0.02}^{+0.02}$	&	$1.25_{-0.15}^{+0.15}$	&	$6.99_{-0.03}^{+0.03}$	&	$0.35_{-0.14}^{+0.14}$	\\

\bottomrule																
\end{tabular}																
\end{adjustbox}																
\label{appendix:F}																	
\end{table*}

\begin{table*}
\centering
\caption{Fe lines gaussian parameters for the dips within the out-of-eclipse observation. All $\sigma$ values were set to 1$\times$10$^{-3}$ keV.}
\begin{adjustbox}{max width=\textwidth}
\begin{tabular}{cccccccc}
\hline\hline
%Out-of-elcipse	&		&		&		&		&		&		&			\\\\
Spectra	&	Orb. Phase	&	Fe\,K$\alpha$ Centroid	&	Fe\,K$\alpha$  Norm	&	\ion{Fe}{xxv} Centroid	&	\ion{Fe}{xxv} Norm	&	\ion{Fe}{xxvi} Centroid &	\ion{Fe}{xxvi} Norm		\\
	&		&	(keV)	&	($\times10^{-3}$ ph cm$^{-2}$ s$^{-1}$)	&	(keV)	&($\times10^{-3}$ ph cm$^{-2}$ s$^{-1}$)	&	(keV)	&	($\times10^{-3}$ ph cm$^{-2}$ s$^{-1}$)	\\
		\midrule
1 	&	 0.35 	&	$6.41_{-0.02}^{+0.03}$	&	$2.5_{-0.4}^{+0.4}$	&	$6.67_{-0.03}^{+0.04}$	&	$1.7_{-0.4}^{+0.4}$	&	$6.97_{-0.03}^{+0.04}$	&	$1.4_{-0.4}^{+0.3}$	\\\\
2 	&	 0.37 	&	$6.41_{-0.02}^{+0.02}$	&	$2.1_{-0.3}^{+0.3}$	&	$6.68_{-0.03}^{+0.02}$	&	$2.4_{-0.3}^{+0.3}$	&	$6.99_{-0.03}^{+0.04}$	&	$1.1_{-0.3}^{+0.3}$	\\\\
3 	&	 0.39 	&	$6.41_{-0.03}^{+0.03}$	&	$2.0_{-0.3}^{+0.3}$	&	$6.66_{-0.04}^{+0.02}$	&	$2.4_{-0.3}^{+0.3}$	&	$6.97_{-0.04}^{+0.04}$	&	$1.2_{-0.3}^{+0.3}$	\\\\
4 	&	 0.41 	&	$6.41_{-0.01}^{+0.02}$	&	$2.50_{-0.24}^{+0.24}$	&	$6.68_{-0.02}^{+0.01}$	&	$2.8_{-0.2}^{+0.3}$	&	$6.97_{-0.02}^{+0.03}$	&	$1.15_{-0.24}^{+0.24}$	\\\\
5 	&	 0.43 	&	$6.41_{-0.02}^{+0.02}$	&	$2.0_{-0.3}^{+0.3}$	&	$6.68_{-0.03}^{+0.02}$	&	$2.4_{-0.3}^{+0.3}$	&	$6.97_{-0.03}^{+0.06}$	&	$0.9_{-0.3}^{+0.3}$	\\\\
6 	&	 0.44 	&	$6.41_{-0.02}^{+0.02}$	&	$2.2_{-0.3}^{+0.3}$	&	$6.68_{-0.02}^{+0.02}$	&	$2.5_{-0.3}^{+0.3}$	&	$6.97_{-0.02}^{+0.06}$	&	$1.0_{-0.3}^{+0.3}$	\\\\
7 	&	 0.46 	&	$6.41_{-0.02}^{+0.02}$	&	$2.2_{-0.4}^{+0.4}$	&	$6.68_{-0.02}^{+0.02}$	&	$2.7_{-0.4}^{+0.4}$	&	$6.99_{-0.02}^{+0.06}$	&	$0.8_{-0.4}^{+0.4}$	\\\\
8 	&	 0.46 	&	$6.41_{-0.02}^{+0.02}$	&	$2.4_{-0.4}^{+0.3}$	&	$6.67_{-0.03}^{+0.03}$	&	$2.0_{-0.4}^{+0.4}$	&	$6.97_{-0.03}^{+0.06}$	&	$0.8_{-0.3}^{+0.3}$	\\\\
9 	&	 0.5 	&	$6.40_{-0.07}^{+0.04}$	&	$1.9_{-0.5}^{+0.5}$	&	$6.66_{-0.07}^{+0.05}$	&	$2.1_{-0.5}^{+0.5}$	&	$6.97_{-0.07}^{+0.20}$	&	$0.3_{-0.3}^{+0.5}$	\\\\
10 	&	 0.52 	&	$6.41_{-0.02}^{+0.03}$	&	$2.2_{-0.4}^{+0.4}$	&	$6.68_{-0.06}^{+0.04}$	&	$2.0_{-0.4}^{+0.4}$	&	$6.97_{-0.06}^{+0.10}$	&	$1.0_{-0.4}^{+0.4}$	\\\\
11 	&	 0.53 	&	$6.41_{-0.03}^{+0.02}$	&	$1.8_{-0.3}^{+0.3}$	&	$6.68_{-0.04}^{+0.03}$	&	$2.1_{-0.3}^{+0.3}$	&	$6.97_{-0.04}^{+0.17}$	&	$0.4_{-0.3}^{+0.3}$	\\\\
12 	&	 0.56 	&	$6.41_{-0.02}^{+0.02}$	&	$2.3_{-0.4}^{+0.3}$	&	$6.68_{-0.05}^{+0.04}$	&	$2.0_{-0.3}^{+0.4}$	&	$6.97_{-0.05}^{+0.05}$	&	$0.5_{-0.3}^{+0.3}$	\\\\
13 	&	 0.58 	&	$6.41_{-0.02}^{+0.03}$	&	$2.7_{-0.4}^{+0.4}$	&	$6.68_{-0.03}^{+0.02}$	&	$2.6_{-0.4}^{+0.4}$	&	$6.99_{-0.03}^{+0.23}$	&	$0.6_{-0.4}^{+0.4}$	\\\\
14 	&	 0.59 	&	$6.41_{-0.04}^{+0.03}$	&	$2.5_{-0.5}^{+0.5}$	&	$6.66_{-0.05}^{+0.06}$	&	$1.8_{-0.6}^{+0.6}$	&	$6.99_{-0.05}^{+0.06}$	&	$1.3_{-0.5}^{+0.5}$	\\\\
15 	&	 0.6 	&	$6.41_{-0.02}^{+0.02}$	&	$2.2_{-0.4}^{+0.4}$	&	$6.67_{-0.03}^{+0.02}$	&	$2.5_{-0.3}^{+0.4}$	&	$6.98_{-0.03}^{+0.07}$	&	$1.1_{-0.3}^{+0.3}$	\\\\
16 	&	 0.62 	&	$6.41_{-0.03}^{+0.02}$	&	$2.2_{-0.4}^{+0.4}$	&	$6.66_{-0.03}^{+0.03}$	&	$2.3_{-0.4}^{+0.4}$	&	$6.99_{-0.03}^{+0.07}$	&	$0.9_{-0.4}^{+0.4}$	\\\\
17 	&	 0.63 	&	$6.40_{-0.02}^{+0.02}$	&	$2.07_{-0.24}^{+0.24}$	&	$6.68_{-0.02}^{+0.02}$	&	$2.12_{-0.24}^{+0.24}$	&	$6.97_{-0.02}^{+0.07}$	&	$0.34_{-0.24}^{+0.24}$	\\\\
18 	&	 0.65 	&	$6.40_{-0.04}^{+0.04}$	&	$2.0_{-0.4}^{+0.4}$	&	$6.68_{-0.05}^{+0.05}$	&	$1.7_{-0.4}^{+0.5}$	&	$6.97_{-0.05}^{+0.07}$	&	$1.1_{-0.4}^{+0.4}$	\\\\
19 	&	 0.66 	&	$6.41_{-0.02}^{+0.02}$	&	$2.2_{-0.3}^{+0.3}$	&	$6.68_{-0.03}^{+0.02}$	&	$1.7_{-0.3}^{+0.3}$	&	$6.99_{-0.03}^{+0.06}$	&	$0.62_{-0.24}^{+0.24}$	\\\\
20 	&	 0.68 	&	$6.41_{-0.03}^{+0.03}$	&	$1.8_{-0.3}^{+0.3}$	&	$6.66_{-0.04}^{+0.04}$	&	$1.8_{-0.3}^{+0.3}$	&	$6.97_{-0.04}^{+0.06}$	&	$0.4_{-0.3}^{+0.3}$	\\\\
21 	&	 0.68 	&	$6.41_{-0.02}^{+0.02}$	&	$2.3_{-0.3}^{+0.3}$	&	$6.66_{-0.03}^{+0.02}$	&	$1.3_{-0.3}^{+0.3}$	&	$6.99_{-0.03}^{+0.06}$	&	$0.7_{-0.3}^{+0.2}$	\\\\
22 	&	 0.69 	&	$6.41_{-0.02}^{+0.03}$	&	$2.4_{-0.2}^{+0.3}$	&	$6.66_{-0.06}^{+0.04}$	&	$0.8_{-0.2}^{+0.3}$	&	$6.97_{-0.06}^{+0.06}$	&	$0.4_{-0.2}^{+0.3}$	\\\\
23 	&	 0.7 	&	$6.41_{-0.02}^{+0.03}$	&	$2.4_{-0.5}^{+0.3}$	&	$6.66_{-0.05}^{+0.07}$	&	$1.1_{-0.5}^{+0.3}$	&	$6.99_{-0.05}^{+0.08}$	&	$1.0_{-0.4}^{+0.3}$	\\\\
24 	&	 0.71 	&	$6.41_{-0.02}^{+0.01}$	&	$2.2_{-0.3}^{+0.2}$	&	$6.66_{-0.02}^{+0.07}$	&	$1.4_{-0.2}^{+0.3}$	&	$6.97_{-0.02}^{+0.09}$	&	$0.71_{-0.24}^{+0.24}$	\\\\
25 	&	 0.72 	&	$6.41_{-0.05}^{+0.03}$	&	$2.6_{-0.4}^{+0.5}$	&	$6.66_{-0.21}^{+0.05}$	&	$0.9_{-0.4}^{+0.5}$	&	$6.97_{-0.21}^{+0.05}$	&	$0.8_{-0.4}^{+0.4}$	\\\\
26 	&	 0.73 	&	$6.41_{-0.02}^{+0.02}$	&	$2.5_{-0.5}^{+0.3}$	&	$6.66_{-0.04}^{+0.03}$	&	$2.1_{-0.4}^{+0.3}$	&	$6.99_{-0.04}^{+0.04}$	&	$0.01_{-0.01}^{+0.23}$	\\\\
27 	&	 0.73 	&	$6.41_{-0.03}^{+0.02}$	&	$2.1_{-0.5}^{+0.5}$	&	$6.68_{-0.05}^{+0.16}$	&	$1.2_{-0.4}^{+0.4}$	&	$7.0_{-0.1}^{+0.3}$	&	$0.5_{-0.4}^{+0.4}$	\\\\
28 	&	 0.75 	&	$6.41_{-0.02}^{+0.03}$	&	$2.0_{-0.3}^{+0.4}$	&	$6.66_{-0.03}^{+0.04}$	&	$1.1_{-0.3}^{+0.5}$	&	$7_{-0}^{+7}$	&	$0.6_{-0.3}^{+0.5}$	\\\\
29 	&	 0.76 	&	$6.41_{-0.02}^{+0.03}$	&	$1.7_{-0.3}^{+0.4}$	&	$6.68_{-0.04}^{+0.06}$	&	$1.3_{-0.2}^{+0.4}$	&	$6.99_{-0.04}^{+0.06}$	&	$0.5_{-0.3}^{+0.4}$	\\\\
30 	&	 0.78 	&	$6.41_{-0.01}^{+0.02}$	&	$2.2_{-0.3}^{+0.3}$	&	$6.68_{-0.06}^{+0.03}$	&	$1.01_{-0.23}^{+0.24}$	&	$6.99_{-0.06}^{+0.04}$	&	$0.6_{-0.3}^{+0.3}$	\\\\
31 	&	 0.79 	&	$6.41_{-0.02}^{+0.02}$	&	$1.97_{-0.24}^{+0.24}$	&	$6.66_{-0.05}^{+0.03}$	&	$0.99_{-0.24}^{+0.24}$	&	$6.99_{-0.05}^{+0.06}$	&	$0.33_{-0.24}^{+0.24}$	\\

\bottomrule																
\end{tabular}																
\end{adjustbox}																
\label{appendix:G}																	
\end{table*}

\begin{table*}
\centering
\caption{Fe lines gaussian parameters for the out-of-dip regions within the out-of-eclipse observation. All $\sigma$ values were set to 1$\times$10$^{-3}$ keV.}
\begin{adjustbox}{max width=\textwidth}
\begin{tabular}{cccccccc}
\hline\hline
%Out-of-elcipse	&		&		&		&		&		&		&			\\\\
Spectra	&	Orb. Phase	&	Fe\,K$\alpha$ Centroid	&	Fe\,K$\alpha$ Norm	&	\ion{Fe}{xxv} Centroid	&	\ion{Fe}{xxv} Norm	&	\ion{Fe}{xxvi} Centroid &	\ion{Fe}{xxvi} Norm		\\
	&		&	(keV)	&	($\times10^{-3}$ ph cm$^{-2}$ s$^{-1}$)	&	(keV)	&($\times10^{-3}$ ph cm$^{-2}$ s$^{-1}$)	&	(keV)	&	($\times10^{-3}$ ph cm$^{-2}$ s$^{-1}$)		\\
		\midrule
1 	&	 0.36 	&	$6.41_{-0.01}^{+0.01}$	&	$2.33_{-0.22}^{+0.18}$	&	$6.68_{-0.02}^{+0.03}$	&	$2.3_{-0.18}^{+0.22}$	&	$6.97_{-0.04}^{+0.07}$	&	$1.0_{-0.2}^{+1.0}$	\\\\
2 	&	 0.38 	&	$6.41_{-0.01}^{+0.02}$	&	$2.6_{-0.3}^{+0.3}$	&	$6.68_{-0.01}^{+0.02}$	&	$2.8_{-0.3}^{+0.3}$	&	$6.99_{-0.03}^{+0.03}$	&	$1.4_{-0.2}^{+1.4}$	\\\\
3 	&	 0.42 	&	$6.41_{-0.01}^{+0.02}$	&	$2.59_{-0.24}^{+0.22}$	&	$6.68_{-0.02}^{+0.02}$	&	$2.47_{-0.23}^{+0.24}$	&	$6.98_{-0.04}^{+0.05}$	&	$1.1_{-0.2}^{+1.1}$	\\\\
4 	&	 0.43 	&	$6.41_{-0.02}^{+0.01}$	&	$2.3_{-0.3}^{+0.3}$	&	$6.67_{-0.02}^{+0.02}$	&	$2.4_{-0.3}^{+0.3}$	&	$6.97_{-0.05}^{+0.04}$	&	$1.0_{-0.3}^{+1.0}$	\\\\
5 	&	 0.45 	&	$6.41_{-0.02}^{+0.02}$	&	$2.2_{-0.3}^{+0.3}$	&	$6.67_{-0.02}^{+0.02}$	&	$2.5_{-0.3}^{+0.3}$	&	$6.99_{-0.08}^{+0.05}$	&	$1.0_{-0.3}^{+1.0}$	\\\\
6 	&	 0.46 	&	$6.41_{-0.03}^{+0.03}$	&	$2.2_{-0.5}^{+0.5}$	&	$6.68_{-0.05}^{+0.05}$	&	$2.6_{-0.5}^{+0.5}$	&	$6.97_{-0.06}^{+0.09}$	&	$1.2_{-0.5}^{+1.2}$	\\\\
7 	&	 0.5 	&	$6.41_{-0.01}^{+0.01}$	&	$2.3_{-0.2}^{+0.2}$	&	$6.68_{-0.01}^{+0.02}$	&	$2.0_{-0.2}^{+0.2}$	&	$6.97_{-0.03}^{+0.03}$	&	$0.6_{-0.2}^{+0.6}$	\\\\
8 	&	 0.52 	&	$6.41_{-0.01}^{+0.01}$	&	$2.38_{-0.24}^{+0.22}$	&	$6.68_{-0.03}^{+0.02}$	&	$2.11_{-0.24}^{+0.24}$	&	$6.97_{-0.06}^{+0.05}$	&	$0.9_{-0.2}^{+0.9}$	\\\\
9 	&	 0.57 	&	$6.41_{-0.02}^{+0.01}$	&	$2.54_{-0.16}^{+0.14}$	&	$6.68_{-0.02}^{+0.01}$	&	$2.24_{-0.14}^{+0.16}$	&	$6.97_{-0.03}^{+0.03}$	&	$0.7_{-0.2}^{+0.7}$	\\\\
10 	&	 0.58 	&	$6.41_{-0.03}^{+0.02}$	&	$2.6_{-0.5}^{+0.4}$	&	$6.67_{-0.05}^{+0.04}$	&	$1.9_{-0.4}^{+0.4}$	&	$6.99_{-0.08}^{+0.09}$	&	$0.8_{-0.4}^{+0.8}$	\\\\
11 	&	 0.59 	&	$6.41_{-0.03}^{+0.02}$	&	$2.5_{-0.4}^{+0.4}$	&	$6.68_{-0.06}^{+0.03}$	&	$2.6_{-0.4}^{+0.4}$	&	$6.97_{-0.04}^{+0.05}$	&	$0.6_{-0.4}^{+0.6}$	\\\\
12 	&	 0.6 	&	$6.41_{-0.01}^{+0.02}$	&	$2.35_{-0.20}^{+0.18}$	&	$6.68_{-0.01}^{+0.01}$	&	$2.45_{-0.18}^{+0.19}$	&	$6.97_{-0.05}^{+0.15}$	&	$0.9_{-0.2}^{+0.9}$	\\\\
13 	&	 0.62 	&	$6.41_{-0.02}^{+0.01}$	&	$2.29_{-0.24}^{+0.23}$	&	$6.67_{-0.03}^{+0.02}$	&	$2.3_{-0.2}^{+0.3}$	&	$6.97_{-0.07}^{+0.05}$	&	$0.7_{-0.2}^{+0.7}$	\\\\
14 	&	 0.64 	&	$6.41_{-0.01}^{+0.03}$	&	$2.22_{-0.24}^{+0.21}$	&	$6.66_{-0.02}^{+0.03}$	&	$2.43_{-0.20}^{+0.22}$	&	$6.97_{-0.03}^{+0.03}$	&	$0.8_{-0.2}^{+0.8}$	\\\\
15 	&	 0.65 	&	$6.41_{-0.01}^{+0.01}$	&	$2.37_{-0.24}^{+0.23}$	&	$6.67_{-0.02}^{+0.02}$	&	$2.10_{-0.22}^{+0.23}$	&	$6.97_{-0.08}^{+0.07}$	&	$0.3_{-0.2}^{+0.3}$	\\\\
16 	&	 0.66 	&	$6.41_{-0.01}^{+0.02}$	&	$2.8_{-0.3}^{+0.3}$	&	$6.68_{-0.03}^{+0.03}$	&	$1.4_{-0.3}^{+0.3}$	&	$6.99_{-0.06}^{+0.07}$	&	$0.7_{-0.3}^{+0.7}$	\\\\
17 	&	 0.67 	&	$6.41_{-0.02}^{+0.04}$	&	$2.1_{-0.4}^{+0.4}$	&	$6.67_{-0.03}^{+0.04}$	&	$2.3_{-0.4}^{+0.4}$	&	$6.97_{-0.08}^{+0.09}$	&	$0.5_{-0.4}^{+0.5}$	\\\\
18 	&	 0.68 	&	$6.41_{-0.01}^{+0.02}$	&	$2.8_{-0.3}^{+0.3}$	&	$6.68_{-0.04}^{+0.04}$	&	$1.2_{-0.3}^{+0.3}$	&	$6.97_{-0.08}^{+0.10}$	&	$0.6_{-0.3}^{+0.6}$	\\\\
19 	&	 0.69 	&	$6.41_{-0.01}^{+0.02}$	&	$2.5_{-0.3}^{+0.3}$	&	$6.66_{-0.04}^{+0.02}$	&	$1.4_{-0.3}^{+0.3}$	&	$6.99_{-0.05}^{+0.06}$	&	$0.5_{-0.3}^{+0.5}$	\\\\
20 	&	 0.7 	&	$6.41_{-0.04}^{+0.02}$	&	$2.2_{-0.5}^{+0.5}$	&	$6.66_{-0.07}^{+0.05}$	&	$1.2_{-0.5}^{+0.5}$	&	$6.97_{-0.06}^{+0.05}$	&	$1.2_{-0.4}^{+1.2}$	\\\\
21 	&	 0.71 	&	$6.41_{-0.02}^{+0.02}$	&	$2.2_{-0.3}^{+0.3}$	&	$6.66_{-0.07}^{+0.02}$	&	$1.4_{-0.3}^{+0.3}$	&	$6.97_{-0.09}^{+0.06}$	&	$0.7_{-0.3}^{+0.5}$	\\\\
22 	&	 0.72 	&	$6.41_{-0.01}^{+0.02}$	&	$2.4_{-0.3}^{+0.5}$	&	$6.67_{-0.03}^{+0.05}$	&	$1.6_{-0.3}^{+0.5}$	&	$6.99_{-0.06}^{+0.05}$	&	$0.3_{-0.3}^{+1.2}$	\\\\
23 	&	 0.72 	&	$6.39_{-0.07}^{+0.04}$	&	$2.3_{-0.8}^{+0.3}$	&	$6.68_{-0.10}^{+0.06}$	&	$1.8_{-0.8}^{+0.3}$	&	$6.98_{-0.23}^{+0.11}$	&	$0.5_{-0.6}^{+0.7}$	\\\\
24 	&	 0.73 	&	$6.41_{-0.01}^{+0.02}$	&	$2.4_{-0.2}^{+0.3}$	&	$6.68_{-0.02}^{+0.02}$	&	$1.7_{-0.2}^{+0.3}$	&	$6.99_{-0.07}^{+0.05}$	&	$0.7_{-0.2}^{+0.3}$	\\\\
25 	&	 0.75 	&	$6.41_{-0.02}^{+0.07}$	&	$2.6_{-0.3}^{+0.9}$	&	$6.67_{-0.06}^{+0.06}$	&	$1.2_{-0.3}^{+0.7}$	&	$6.97_{-0.05}^{+0.1}$	&	$0.7_{-0.3}^{+0.5}$	\\\\
26 	&	 0.76 	&	$6.41_{-0.02}^{+0.02}$	&	$2.4_{-0.3}^{+0.2}$	&	$6.68_{-0.06}^{+0.03}$	&	$1.6_{-0.3}^{+0.2}$	&	$6.97_{-0.05}^{+0.06}$	&	$0.5_{-0.3}^{+0.7}$	\\\\
27 	&	 0.77 	&	$6.41_{-0.01}^{+0.02}$	&	$2.1_{-0.2}^{+0.3}$	&	$6.68_{-0.03}^{+0.04}$	&	$1.2_{-0.2}^{+0.3}$	&	$6.97_{-0.05}^{+0.05}$	&	$0.4_{-0.2}^{+0.7}$	\\

\bottomrule																
\end{tabular}																
\end{adjustbox}																
\label{appendix:H}																	
\end{table*}

%%%%%%%%%%%%%%%%%%%%%%%%%%%%%%%%%%%%%%%%%%%%%%%%%%

% Don't change these lines
\bsp	% typesetting comment
\label{lastpage}
\end{document}